\documentclass[superscriptaddress, aps, pre, reprint, floatfix]{revtex4-2}
\usepackage{graphicx,hyperref,color,upgreek}
\usepackage{amsmath}
\usepackage{latexsym}
\usepackage{float}
\usepackage{amssymb}
\usepackage{multirow}
\usepackage{graphicx}
\usepackage{textcomp}
\usepackage{hyperref}
\usepackage{array}
\usepackage{xcolor}
\usepackage{hyperref}
\usepackage{xkeyval,xcolor}
\usepackage{xcite}
\graphicspath{{Main/figures/},{Main/}}
\usepackage[caption=false]{subfig}
\usepackage{dcolumn}
\usepackage{bm}
\usepackage{xr}
\usepackage{tikz}
\usepackage{tabularx}
\makeatletter
\newcommand*{\addFileDependency}[1]{
	\typeout{(#1)}
	\@addtofilelist{#1}
	\IfFileExists{#1}{}{\typeout{No file #1.}}
}

\makeatother

\newcommand*{\myexternaldocument}[1]{%
	\externaldocument{#1}%
	\addFileDependency{#1.tex}%
	\addFileDependency{#1.aux}%
}

\myexternaldocument{main_ESI}

\def \figwidth {9cm}
\def \newfigwidth {8cm}

\begin{document}

\preprint{APS/123-QED}

\title{Wrapping nonspherical vesicles at bio-membranes}

\author{Ajit Kumar Sahu} 
\affiliation{Department of Physics, School of Basic Sciences, Indian Institute of Technology Bhubaneswar, Jatni, Odisha-752050, India}%

\author{Rajkumar Malik}
\affiliation{Department of Physics, School of Basic Sciences, Indian Institute of Technology Bhubaneswar, Jatni, Odisha-752050, India}%

\author{Jiarul Midya}
\email{jmidya@iitbbs.ac.in}
\affiliation{Department of Physics, School of Basic Sciences, Indian Institute of Technology Bhubaneswar, Jatni, Odisha-752050, India}%

\date{\today}

\begin{abstract}
The wrapping of particles and vesicles by lipid bilayer membranes is a fundamental process in cellular transport and targeted drug delivery. Here, we investigate the wrapping behavior of nonspherical vesicles, such as ellipsoidal, prolate, oblate, and stomatocytes, by systematically varying the bending rigidity of the vesicle membrane and the tension of the planar membrane. Using the Helfrich Hamiltonian, triangulated membrane models, and energy minimization techniques, we predict multiple stable wrapping states and identify the conditions for their coexistence. Our results demonstrate that softer vesicles bind more easily to planar membranes; however, achieving complete wrapping requires significantly higher adhesion strengths compared to rigid particles. As membrane tension increases, deep-wrapped states disappear at a triple point where shallow-wrapped, deep-wrapped, and complete-wrapped states coexist. The coordinates of the triple point are highly sensitive to the vesicle shape and stiffness. For stomatocytes, increasing stiffness shifts the triple point to higher adhesion strengths and membrane tensions, while for oblates it shifts to lower values, influenced by shape changes during wrapping. Oblate shapes are preferred in shallow-wrapped states and stomatocytes in deep-wrapped states. In contrast to hard particles, where optimal adhesion strength for complete wrapping occurs at tensionless membranes, complete wrapping of soft vesicles requires finite membrane tension for optimal adhesion strength. These findings provide new insights into the interplay between vesicle deformability, shape, and membrane properties, advancing our understanding of endocytosis and the design of advanced biomimetic delivery systems.
\end{abstract}

\maketitle

\section{Introduction}
Cellular transport of information and materials, a fundamental process that ensures the survival and functionality of cells, is typically facilitated by the exchange of particles across lipid membranes \cite{POMORSKI2001139, Sprong2001504}. The translocation across lipid-bilayer membranes depends on the size, elasticity, shape, and surface functionalization of the particles\cite{dasgupta2017nano}. Particles smaller than the thickness of the membrane can translocate, causing minimal membrane deformation \cite{doi:10.1021/nn3028858, MacroLetter2017_Sommer}, while particles larger than a few nanometers translocate through endocytosis \cite{endocytosis_acsnano2020, endocytosis_pnas2010}. Passive endocytosis is a multistep process that begins with the attachment of the particle to the membrane, followed by wrapping and finally detachment from the membrane \cite{endocytosis}. The particle wrapping process, a key aspect of endocytosis, is ubiquitous for both living and nonliving particles, such as drug-carrying nanoparticles \cite{kube2017fusogenic}, viruses \cite{wiegand2020forces}, and parasite invasion into host cells\cite{pmid24988340}. Membrane tension, regulated by cells through remodeling, plays a critical role in endocytosis: high tension slows the internalization process, while low tension facilitates membrane invagination and efficient particle uptake \cite{Qingfen_NanoLett2020, PNAS_Gauthier_2013, PhysRevE_2006_Turner}. Understanding the particle wrapping process can help us optimizing biomedical applications, such as targeted drug delivery using engineered nanoparticles \cite{pmid33277608,de2008drug,barbe2004silica}.

The wrapping of hard spherical particles at lipid membranes has been extensively investigated in experimental and theoretical studies \cite{van2016lipid,raatz2014cooperative,deserno2003wrapping,lipowsky1998vesicles,ACSNano_2024_Rao}. In general, for the wrapping of hard particles, the size and shape of the particle, the bending rigidity and tension of the membrane, and the adhesion strength of the particle-membrane have been shown to primarily control the process \cite{ NanoLetters2023_Kraft, doi:10.1021/nl403949h, jiang2008nanoparticle,C3SM50351H, PhysRevE.71.061902,doi:10.1080/00018739700101488, bahrami2013orientational,pmid35671377}. In addition, the initial orientation of the anisotropic hard nanoparticles relative to the adhering membranes influences their wrapping behavior. For example, the wrapping of ellipsoidal \cite{doi:10.1021/nl403949h,huang2013role} and spherocylindrical \cite{ SoftMatter2022_Laradji} nanoparticles starts with an orientation in which the major axis of the particle is parallel to the membrane surface. After half-wrapping, the orientation of the particle changes, and the major axis becomes perpendicular to the membrane surface. Unlike hard particles, soft particles can deform depending on external constraints \cite{doi:10.1021/acsnano.2c05801,yi2011cellular}; the deformability plays an important role in the wrapping process, in addition to the factors that apply to hard particles.  

Many biologically relevant particles have anisotropic or nonspherical shapes \cite{pmid11347712, 10.1371/journal.ppat.1000394}. For example, malaria parasites are typically egg-shaped \cite{pmid24988340}, mature virions can be ellipsoidal or brick-shaped \cite{CONDIT200631, Nature_Liu2024}, and vesicular stomatitis viruses (VSV) have bullet shapes \cite{Science2010_bulletVirus}. Simulations and experiments show that for ellipsoidal particles the stability of the partial-wrapped over non-wrapped and complete-wrapped states decreases when the bending rigidity of the membrane increases \cite{liu2023wrapping}. Living particles also often have the ability to change shape and deformability throughout their life cycles. For example, murine leukemia virus (MLV) and human immunodeficiency virus (HIV) adjust their deformability through internal structural changes, significantly affecting their entry into host cells\cite{pang2013virion, HIV_BioPhys2006, pmid17620615}. The mature HIV viral particles are less stiff compared to the immature HIV viral particles, which hinders their complete wrapping at the host cells \cite{HIV_BioPhys2006}.

A wide range of synthetic deformable particles with various architectures and customizable mechanical properties can be engineered for the targeted delivery of drugs, including star polymers \cite{likos1998star,ren2016star}, microgels \cite{brugnoni2018swelling,plamper2017functional}, dendrimers \cite{chen2023cargo,boas2004dendrimers}, polymer-grafted nanoparticles \cite{midya2019disentangling,vlassopoulos2010polymers}, and vesicles \cite{hoffmann2018dynamics,kube2017fusogenic,doi:10.1080/00018739700101488}. The deformability of these particles can be tuned through control parameters that change their molecular architecture or building blocks. For example, the deformability of microgels can be adjusted by changing the cross-link density and electric charge \cite{hofken2024real,strauch2023ionisation}, of polymer-grafted nanoparticles by altering the density and length of grafted polymers \cite{midya2019disentangling}, and of unilamellar fluid vesicles by modifying the bending rigidity and the osmotic pressure difference \cite{vorselen2017competition}. A particularly versatile and well-characterized class of deformable particles is unilamellar vesicles \cite{doi:10.1080/00018739700101488}. Free vesicles with fixed membrane areas, variable volumes, and symmetric lipid bilayer membranes assume a spherical shape and may change volumes significantly during wrapping. Nonspherical vesicles with volumes smaller than the maximal volume that can be enclosed by the vesicle membrane and potentially even asymmetric monolayer composition can have a complete zoo of shapes \cite{Seifert01021997}.   

In this work, we investigate the wrapping of single nonspherical vesicles with initial ellipsoidal, prolate, oblate, and stomatocytic shapes at planar membranes by systematically varying the relative stiffness of the vesicle membranes and the tension of the planar membranes. Using triangulated membranes and energy minimization techniques, we predict energetically stable wrapping states. The systematic increase in adhesion strength leads to transitions of the vesicles from non-wrapped to complete-wrapped states through intermediate shallow-wrapped and deep-wrapped states. Softer vesicles undergo both shape and orientation changes during wrapping, whereas stiffer vesicles exhibit only orientation changes. We calculate wrapping diagrams showing the energetically stable states for various membrane bending rigidities and tensions. As membrane tension increases, the deep-wrapped state becomes destabilized and eventually vanishes at a triple point, where shallow-, deep-, and complete-wrapped states coexist.  Beyond the triple point, vesicles exhibit a discontinuous transition from the shallow-wrapped state to the complete-wrapped state. For initial stomatocytes, the triple point shifts to higher membrane tensions and adhesion strengths as the stiffness of the vesicles increases, while the trend is the opposite for initially oblate ones. In addition, we show that the optimal adhesion strength for the complete wrapping of soft oblate, prolate, and ellipsoidal vesicles occurs at a non-zero membrane tension, unlike rigid particles, which require zero membrane tension. This behavior is related to the shape change of the vesicles during wrapping. Our findings indicate that the initially stomatocyte, oblate and prolate vesicles preferred to be oblate in the shallow-wrapped states and stomatocyte in the deep-wrapped states.

The rest of the manuscript is organized as follows. In Sec.~\ref{sec:model}, we introduce the model and the numerical calculation techniques. In Sects.~\ref{sec:states} and \ref{sec:shapes}, we show the deformation energy landscapes along with the typical vesicle shapes and characterize the vesicle shapes at various wrapping stages. In Sec.~\ref{sec:vesstiffness}, we study the effect of vesicle deformability and in Sec.~\ref{sec:memtension} of membrane tension on the stability of the wrapping states. Finally, in Sec.~\ref{sec:conclusions}, we summarize and conclude our study.

\section{\label{sec:model}Model and methods}
\begin{figure}[t]
    \centering
    \includegraphics[width=\figwidth]{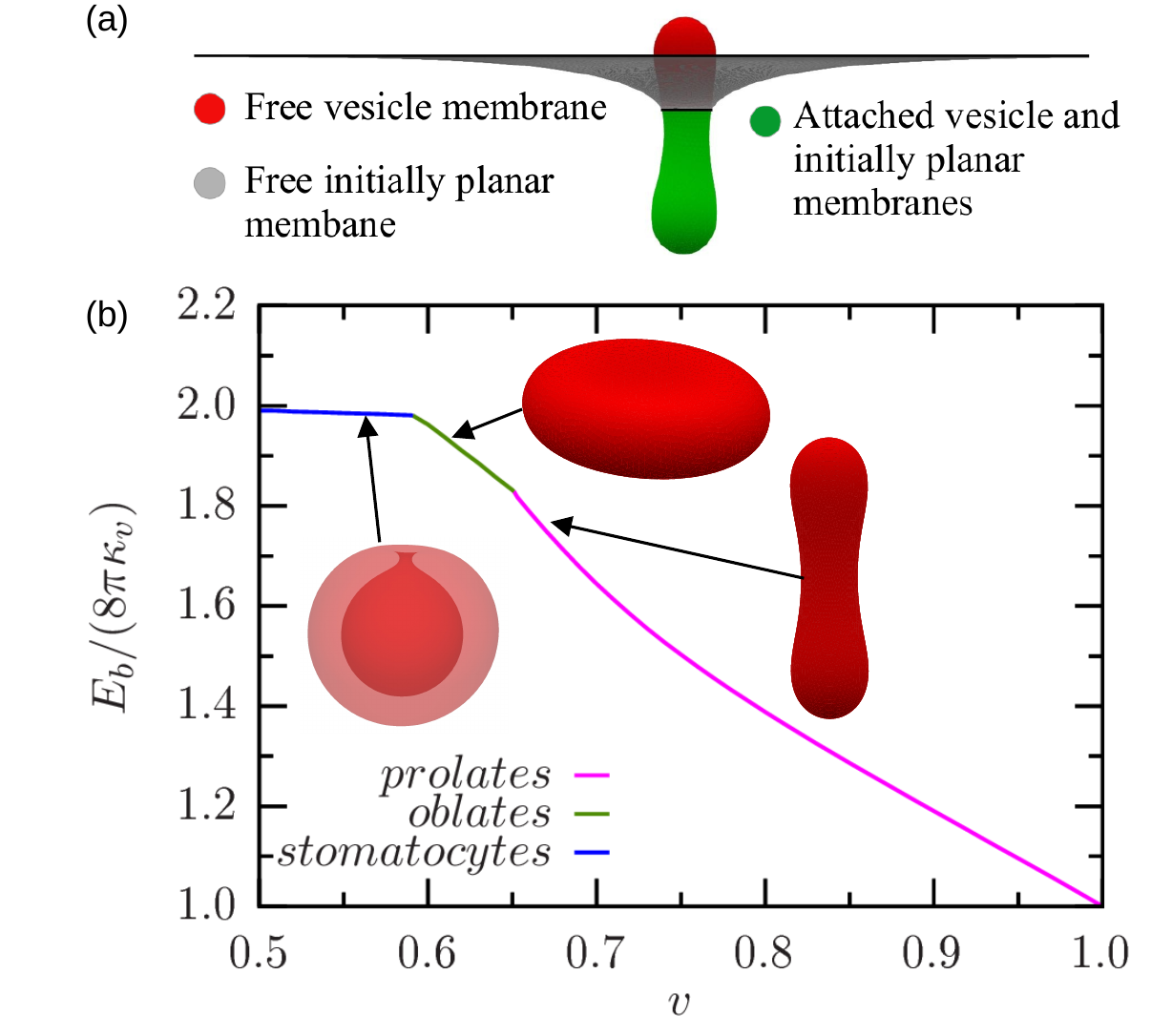}
    \caption{(a) Schematic representation of a partial-wrapped prolate vesicle at a planar membrane. (b) Bending energy of vesicles as a function of reduced volume.}
    \label{fig:fw_figure+VesiShape}
\end{figure}

The wrapping of vesicles at planar membranes is studied using a continuum membrane model, where the total energy of the system is calculated based on the Helfrich Hamiltonian \cite{Helfrich+1973+693+703,deuling:jpa-00208531},
\begin{multline}
   E=\int_{A_{\rm p}}dS \, \left[ 2\kappa_{\rm p} H^2+\sigma \right]+2\kappa_{\rm v} \int_{A_{\rm v}} dS \, H^2-w\int_{A_{\rm ad}}dS \\ +\sigma_{\rm v} A_{\rm v} + p_{\rm v} V_{\rm v} \, . 
   \label{eq:Helfrich_Hamiltonian}
\end{multline}
Here, $H=(c_1+c_2)/2$ is the mean curvature, $c_1$ and $c_2$ the principal curvatures, $A_{\rm v}$ the total area of the vesicle, and $A_{\rm p}$ the area of the planar membrane. The adhered area $A_{\rm ad}$ is the area over which the vesicle and the planar membrane are attached to each other with adhesion strength $w$, see Fig.~\ref{fig:fw_figure+VesiShape}(a). The bending rigidities of the vesicle and planar membranes are $\kappa_{\rm v}$ and $\kappa_{\rm p}$, respectively, and the tension of the planar membrane is $\sigma$. The membrane area $A_{\rm v}$ and the volume $V_{\rm v}$ of the vesicle are conserved using the Lagrange multipliers $\sigma_{\rm v}$ and $p_{\rm v}$, respectively. The tension energy of the vesicle is omitted because its membrane area remains constant throughout the wrapping process.

The nonspherical shapes of the vesicles are characterized by reduced volume $v = V_{\rm v}/V_{\rm s}$, see Fig.~\ref{fig:fw_figure+VesiShape}(b), and Fig.~S1 in the SI. Here, $V_{\rm v}$ is the actual volume of the vesicle with area $A_{\rm v}$, and $V_{\rm s}$ is the volume of a spherical vesicle with the same membrane area. The maximal value, $v = 1$, corresponds to a spherical vesicle.  The values of Lagrange multipliers $\sigma_{v}$ and $p_v$ depend on the vesicle shape; sudden jumps in their values indicate shape transitions; see Fig.~S2 in the SI. The stable states for vesicles with $0.6515 \le v < 1$ are prolate, with $0.5915 \le v \le 0.6515$ are oblate, and with $v \le 0.5915$ are stomatocytes. We refer to vesicles with $v = 0.95$ as ellipsoidal vesicles \cite{gozdz2007deformations}. The wrapping fraction $f_{\rm w}={A_{\rm ad}}/{A_{\rm v}}$ represents the fraction of the vesicle area that is attached to the planar membrane. The value of $f_{\rm w}$ varies in the range 0 to 1, where $f_{\rm w}=0$ corresponds to the non-wrapped (NW) state and $f_{\rm w}=1$ to the complete-wrapped (CW) state. The shallow-wrapped (SW) state is defined in the range $0 < f_{\rm w} \le 0.5$, and the deep-wrapped (DW) state in the range $0.5 < f_{\rm w} < 1$. In this work, we focus in particular on the wrapping behavior of four vesicle shapes: stomatocyte ($v=0.55$), oblate ($v=0.63$), prolate ($v=0.66$) and ellipsoidal ($v=0.95$).  The selection of vesicle shapes at the phase boundaries between the stomatocyte, oblate, and prolate can result in complex wrapping behaviors compared to prolate vesicles with reduced volumes $v>0.7$, as previously reported in the literature \cite{doi:10.1021/acsnano.2c05801}.

All integrals in Eq.~(\ref{eq:Helfrich_Hamiltonian}) are discretized using triangulated surfaces composed of vertices, edges, and facets \cite{Pezeshkian2020,gompper_triangulated-surface_2004}. We initialize the energy and shape calculations with a few large triangles and then iteratively minimize the energy and refine the triangulation until the desired accuracy in energy is achieved; see Fig.~3S in the SI. The iterative procedure is important to avoid the need for major shape changes of finely discretized membranes, which would require a high number of minimization steps. Small triangles are used where the membrane is highly curved, and large triangles are used where it is almost planar. In addition, the edges are regularly swapped to maintain the fluidity of the membranes \cite{kroll1992conformation}, and vertex-averaging steps ensure that neighboring triangles have similar areas. The numerical calculations are performed with the help of the freely available software Surface Evolver \cite{Brakke1992}. 

The total energy of the system can be expressed in terms of reduced units as 
\begin{equation}
    \begin{split}
       \tilde{E}=\frac{E}{\pi \kappa_{\rm p}}
        = \frac{2}{\pi A_{\rm v}} \left[\int_{A_{\rm p}} dS \, \left[ A_{\rm v} H^2 +\pi \tilde{\sigma} \right] + \kappa_{\rm r} A_{\rm v} \int_{A_{\rm v}} dS \,{ H^2}\right]\\ -\tilde{w}f_{\rm w}+\tilde{\sigma_{\rm v}}A_v+\tilde{p_{\rm v}}V_v 
    \end{split}
\end{equation}
where $\tilde{\sigma}=\sigma A_v /(2\pi \kappa_p)$ is the reduced membrane tension and $\tilde{w}=wA_v/(\pi \kappa_p)$ the reduced adhesion strength. The bending rigidity ratio $\kappa_{\rm r} = \kappa_{\rm v}/\kappa_{\rm p}$ characterizes the relative deformability of the vesicle with respect to the planar membrane. Further we can define reduced deformation energy as 
\begin{equation}
\begin{split}
    \Delta\tilde{E} &= \Delta\tilde{E}_{\rm b,v}+\Delta\tilde{E}_{\rm b, p}+\Delta\tilde{E}_{\rm s, p}+\Delta\tilde{E}_{\rm w}\\
    &=\tilde{E}(f_{\rm w})-\tilde{E}(f_{\rm w} = 0).
    \label{eq:reduced_defEnergy_diff}
\end{split}
\end{equation}
Here, $\Delta\tilde{E}_{\rm b,v}$, $\Delta\tilde{E}_{\rm b,p}$, $\Delta\tilde{E}_{\rm s,p}$, and $\Delta\tilde{E}_{\rm w}$ represent changes in the bending energy of the vesicle membrane, the bending energy of the planar membrane, the surface energy of the planar membrane, and the adhesion energy, respectively. The deformation energy is first calculated as a function of the wrapping fraction without adhesion energy ($\tilde{w}=0$) using Eq. (\ref{eq:Helfrich_Hamiltonian}). The data is then fitted with a piecewise fourth-order polynomial for the shallow-wrapped and the deep-wrapped states. Then, the adhesion energy is subsequently added to the fitted curves to identify stable and unstable states.

\section{Results and discussion}
\subsection{\label{sec:states} Wrapping states and deformation energies}
\begin{figure*}[hbt!]
    \centering
    \includegraphics[width=2.0\columnwidth]{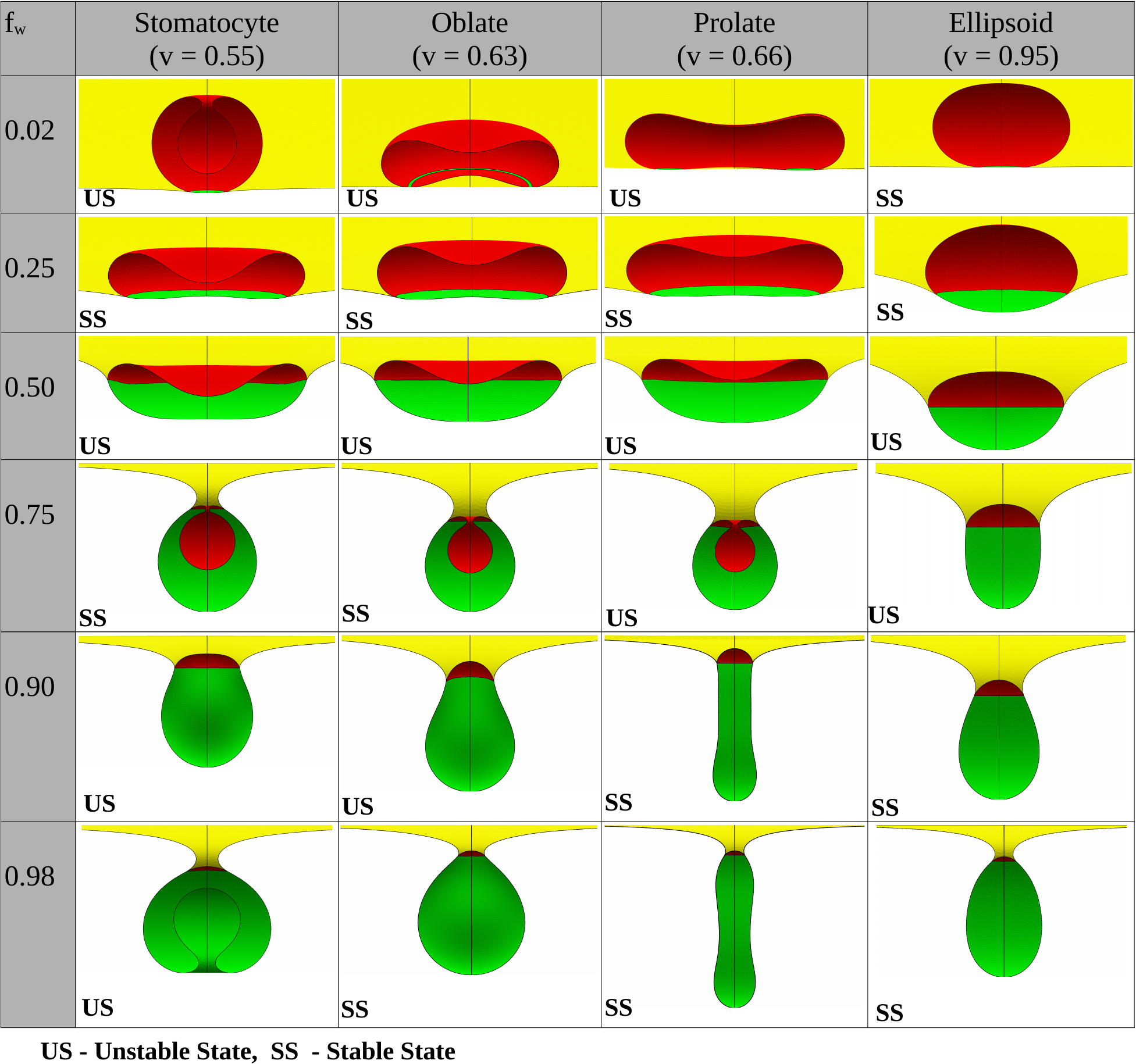}
    \caption{Cross-sectional views of partial-wrapped nonspherical vesicles at various wrapping fractions $f_{\rm w}$ for $\kappa_{\rm r} = 1$ and $\tilde{\sigma}= 0.5 $ along with the indication of stable states (SS) and unstable states (US). Red, green, and yellow colors represent the unattached and attached membrane of the vesicle and the free planar membrane, respectively.}
    \label{fig:snapshots}
\end{figure*}

Figure~\ref{fig:snapshots} illustrates the wrapping states of nonspherical vesicles at planar membranes for various wrapping fractions, $f_{\rm w}$. The wrapping process is consistently initiated in regions of the vesicle with the lowest curvature. For a stomatocyte ($v = 0.55$), initial attachment to the planar membrane occurs in the "rocket" orientation, where the major axis of the vesicle is perpendicular to the planar membrane. As the wrapping progresses, the vesicle deforms to an oblate at $f_{\rm w} \simeq 0.07$ to minimize the deformation energy, and remains oblate within the range $0.07 \leq f_{\rm w} \leq 0.5$. For $f_{\rm w} > 0.5$, the vesicle reverts to its initial stomatocytic shape, persisting in this configuration until the wrapping of the outer sphere of the stomatocyte is completed. At $f_{\rm w} \gtrsim 0.75$, the vesicle transforms into a vertical oblate, avoiding the high bending energy costs of the planar membrane associated with the wrapping of the inner sphere of the stomatocyte. Finally, at $f_{\rm w} > 0.9$, the vesicle returns to an inverted stomatocyte. Upon complete wrapping, both the vesicle and the planar membrane recover their original shapes.

\begin{figure*}[hbt!]
    \centering
    \includegraphics[width=2.0\columnwidth]{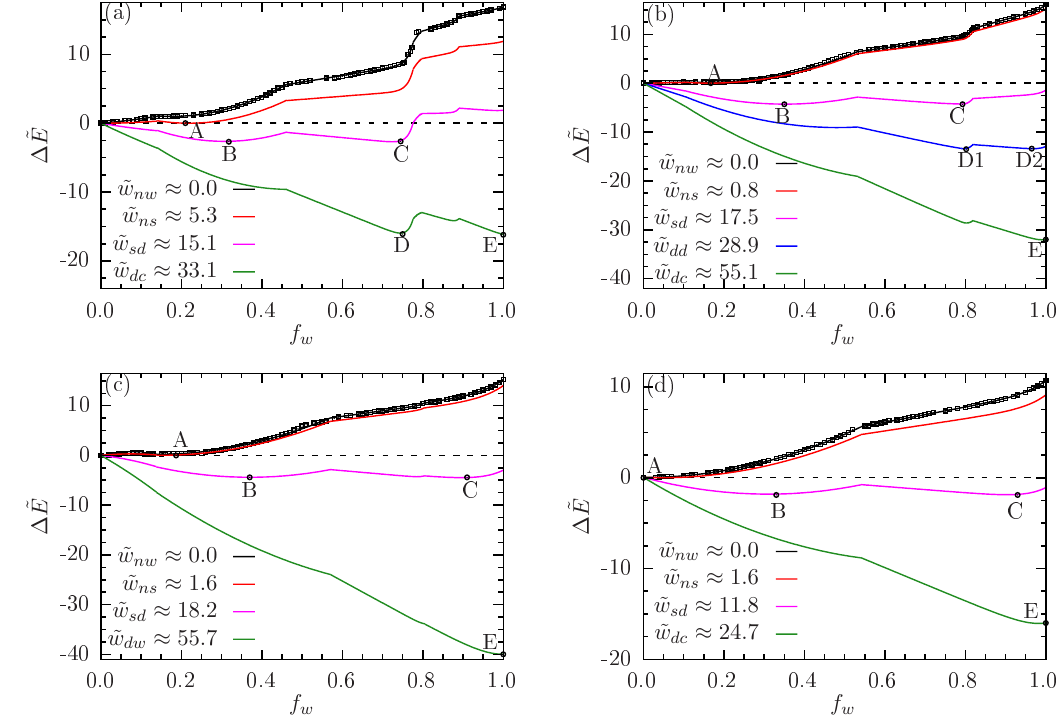}
    \caption{Deformation energy landscapes of nonspherical vesicles at planar membranes as functions of wrapping fraction $f_{\rm w}$ for vesicle with reduced volumes (a) $v= 0.55$ (stomatocyte), (b) $v =0 .63$ (oblate), (c) $v = 0.66$ (prolate), and (d) $v =0.95$ (ellipsoid), at bending rigidity ratio $\kappa_{\rm r} = 1$ and reduced membrane tension $\tilde{\sigma}= 0.5$. The minimal energy states are labeled alphabetically, with A being the wrapping fraction of the shallow-wrapped state for the binding transition, B the wrapping fraction of the shallow-wrapped state, and C the wrapping fraction of the deep-wrapped state for the transition from shallow- to deep-wrapped; D indicates the wrapping fraction of the deep-wrapped state for the envelopment transition; D1 and D2 are wrapping fractions for the deep-wrapped to deep-wrapped transition, and E indicates the complete-wrapped state.}
    \label{fig:deformationEnergy_vs_fw}
\end{figure*}
\begin{figure}[hbt!]
    \centering
    \includegraphics[width=\newfigwidth]{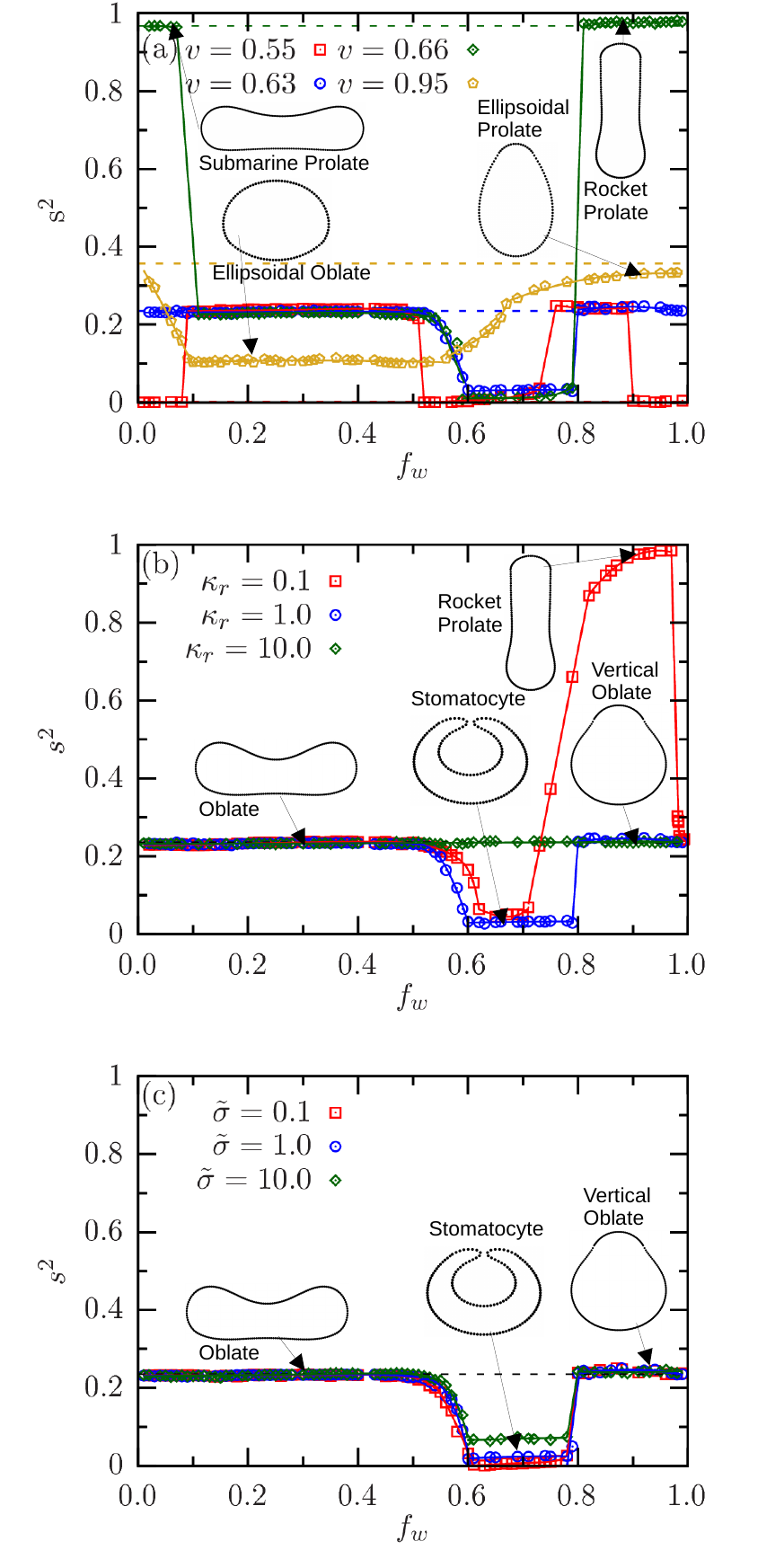}
    \caption{Asphericity variation with wrapping fraction $f_{\rm w}$. (a) Asphericity for vesicles with various $v$ at $\kappa_{\rm r} = 1$ and $\tilde{\sigma} = 0.5$. (b) Asphericity for various $\kappa_{\rm r}$ for $v = 0.63$ at $\tilde{\sigma} = 0.5$.(c) Asphericity for various $\tilde{\sigma}$ for $v = 0.63$ at $\kappa_{\rm r} = 1$. The dashed lines indicate the asphericities of the corresponding free vesicles.}
    \label{fig:aspericity}
\end{figure}

For an oblate vesicle ($v = 0.63$), the initial attachment to the planar membrane occurs in a ring-shaped patch with two contact lines, as shown in Fig.~\ref{fig:snapshots}. As the wrapping fraction $f_{\rm w}$ increases, the radius of the inner contact line decreases and vanishes at $f_{\rm w} \approx 0.1$, forming a circular patch with a single contact line, and remains oblate in the range of wrapping fractions $0.1 < f_{\rm w} \lesssim 0.5$. At higher $f_{\rm w}$ values, the planar membrane comes into contact with the rim of the oblate, resulting in an increased bending energy cost. This energy cost induces a shape change of the vesicle from oblate to stomatocyte, which persists for $0.5 < f_{\rm w} \lesssim 0.79$. As $f_{\rm w}$ increases further, the planar membrane needs to bend more to wrap the inner sphere of the stomatocyte, leading to a sharp increase in bending energy. To minimize the overall energy cost, the vesicle compromises its shape and reverts to a vertical oblate for $f_{\rm w} > 0.79$, maintaining this shape until it is completely wrapped. Similar shape and orientation changes are observed for prolate ($v = 0.66$) and ellipsoidal ($v = 0.95$) vesicles, as illustrated in Fig.~\ref{fig:snapshots}.
\begin{figure*}[hbt!]
    \centering
    \includegraphics[width=2.0\columnwidth]{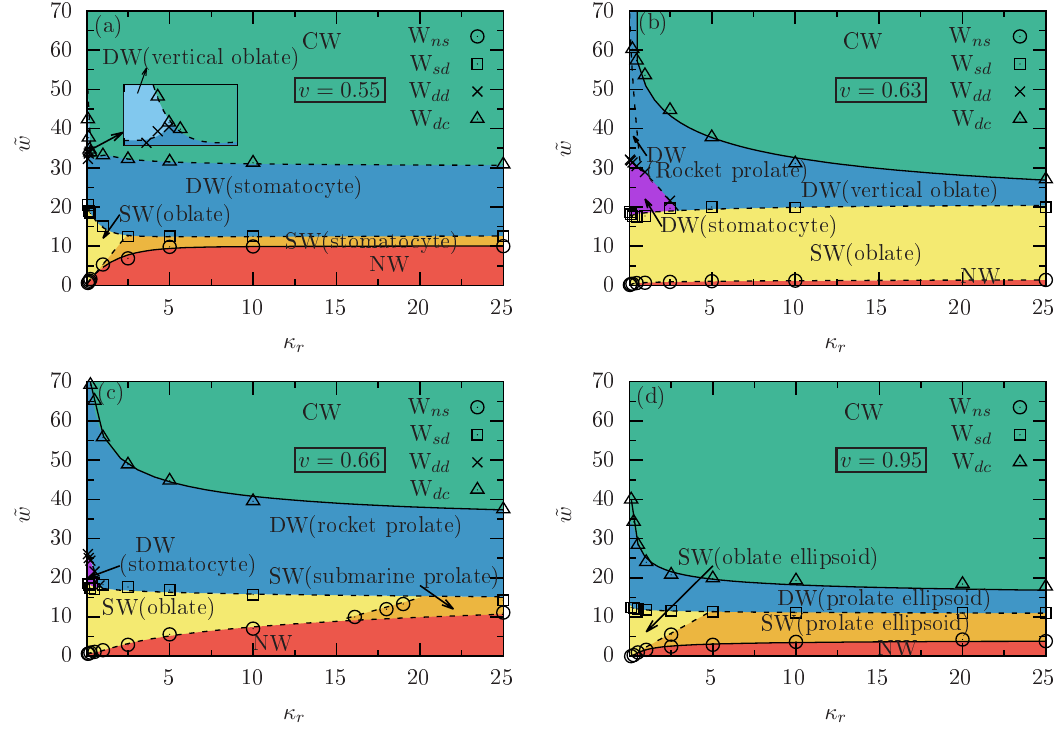}
    \caption{Wrapping diagrams in the $\kappa_{\rm r}$-$\tilde{w}$ plane at $\tilde{\sigma} = 0.5$ show stable states: non-wrapped (NW), shallow-wrapped (SW), deep-wrapped (DW), and complete-wrapped (CW) for vesicles with reduced volumes: (a) $v = 0.55$ (stomatocyte), (b) $v = 0.63$ (oblate), (c) $v = 0.66$ (prolate), and (d) $v = 0.95$ (ellipsoidal). Transitions $W_{\rm ns}$, $W_{\rm sd}$, $W_{\rm dd}$, and $W_{\rm dc}$ indicate NW to SW, SW to DW, DW to DW, and DW to CW transitions, respectively. Solid and dashed lines represent continuous and discontinuous transitions.}
    \label{fig:phasediagram_w_vs_kr}
\end{figure*}

The reduced deformation energy $\Delta \tilde {E}$, calculated using Eqs.~(\ref{eq:Helfrich_Hamiltonian})-(\ref{eq:reduced_defEnergy_diff}), as a function of the wrapping fraction $f_{\rm w}$ for various reduced volumes of vesicles ($v=0.55$, $0.63$, $0.66$ and $0.95$) at a fixed relative stiffness $\kappa_{\rm r} = \kappa_{\rm v}/\kappa_{\rm p} = 1$ and membrane tension $\tilde{\sigma} = 0.5$ is shown in Figs.~\ref{fig:deformationEnergy_vs_fw}(a)-(d). For a stomatocyte ($v=0.55$), $\Delta \tilde{E}$ increases with $f_{\rm w}$ in the absence of adhesion energy ($\tilde{w} = 0$), as shown in Fig.~\ref{fig:deformationEnergy_vs_fw}(a). The energy minimum at $f_{\rm w} = 0$ suggests that the non-wrapped (NW) state is the stable configuration. However, for adhesion strength $\tilde{w} \simeq 5.3$, a local minimum with the same energy is found at $f_{\rm w} \simeq 0.21$ (point `A'). At this adhesion strength, the vesicle undergoes a discontinuous binding transition $W_{\rm ns}$ from a non-wrapped (NW) to a shallow-wrapped (SW) state; stable oblate states can be found for $0.21 \lesssim f_{\rm w} \lesssim 0.32$. At adhesion strength $\tilde{w}~\simeq 15.1$, two energy minima at $f_{\rm w} \simeq 0.32$ (point `B') and $f_{\rm w} \simeq 0.74$ (point `C') have the same height, which corresponds to the discontinuous transition from shallow- to deep-wrapped $W_{\rm sd}$ where oblate and stomatocyte coexist. At even higher adhesion strength, $\tilde {w}\simeq 33.1$, the deep-wrapped stomatocytic state (point `D') coexists with the complete-wrapped (CW) state at $f_{\rm w}=1$ (point `E'); the envelopment transition $W_{\rm dc}$ is also discontinuous.

For an oblate ($v = 0.63$), the reduced deformation energy $\Delta \tilde{E}$ as a function of the wrapping fraction $f_{\rm w}$ is illustrated in Fig.~\ref{fig:deformationEnergy_vs_fw}(b). The first minimum in $\Delta \tilde{E}$, indicating that the binding transition $W_{\rm ns}$ occurs at the adhesion strength $\tilde{w} \simeq 0.8$ and the wrapping fraction $f_{\rm w} \simeq 0.16$. The states within the range $0 \leq f_{\rm w} \lesssim 0.17$ with rim patches, having two contact lines, are on the energy barrier, while the oblate states in the range $0.17 \lesssim f_{\rm w} \lesssim 0.34$ with a single contact line patches are stable. The shallow-to-deep-wrapped transition $W_{\rm sd}$ occurs at adhesion strength $\tilde{w} \simeq 17.5$, where the oblate vesicle at $f_{\rm w} \simeq 0.34$ (point `B') coexists with a stomatocyte at $f_{\rm w} \simeq 0.78$ (point `C'). Within the deep-wrapped (DW) regime, a deep-wrapped (${\rm D}_1$) to deep-wrapped (${\rm D}_2$) transition $W_{\rm dd}$ is observed for $\tilde{w} \simeq 28.9$, where stomatocytes at $f_{\rm w} \simeq 0.8$ (point `D1') coexist with vertical oblates at $f_{\rm w} \simeq 0.96$ (point `D2'). Finally, a continuous envelopment transition $W_{\rm dc}$ occurs with an adhesion strength of $\tilde{w} \simeq 55.1$.

For prolate ($v = 0.66$) and ellipsoidal ($v = 0.95$) vesicles, the deformation energy $\Delta \tilde{E}$ as a function of $f_{\rm w}$ is shown in Figs.~\ref{fig:deformationEnergy_vs_fw}(c) and (d), respectively. The binding transition for prolates is discontinuous, causing a shape transition from prolate to oblate. In contrast, ellipsoidal vesicles ($v = 0.95$) undergo a continuous binding transition, initially attached to the membrane in a submarine orientation, the major axis being parallel to the planar membranes. Both prolate and ellipsoidal vesicles exhibit a discontinuous shallow-to-deep wrapped transition that is accompanied by a shape and orientation change from submarine to rocket, with a rotation of the major axis by 90\textdegree, as shown in Fig.~\ref{fig:snapshots}. The envelopment transition is continuous for both the prolate and ellipsoidal vesicles.

\begin{figure*}[hbt!]
    \centering
    \includegraphics[width=2.0\columnwidth]{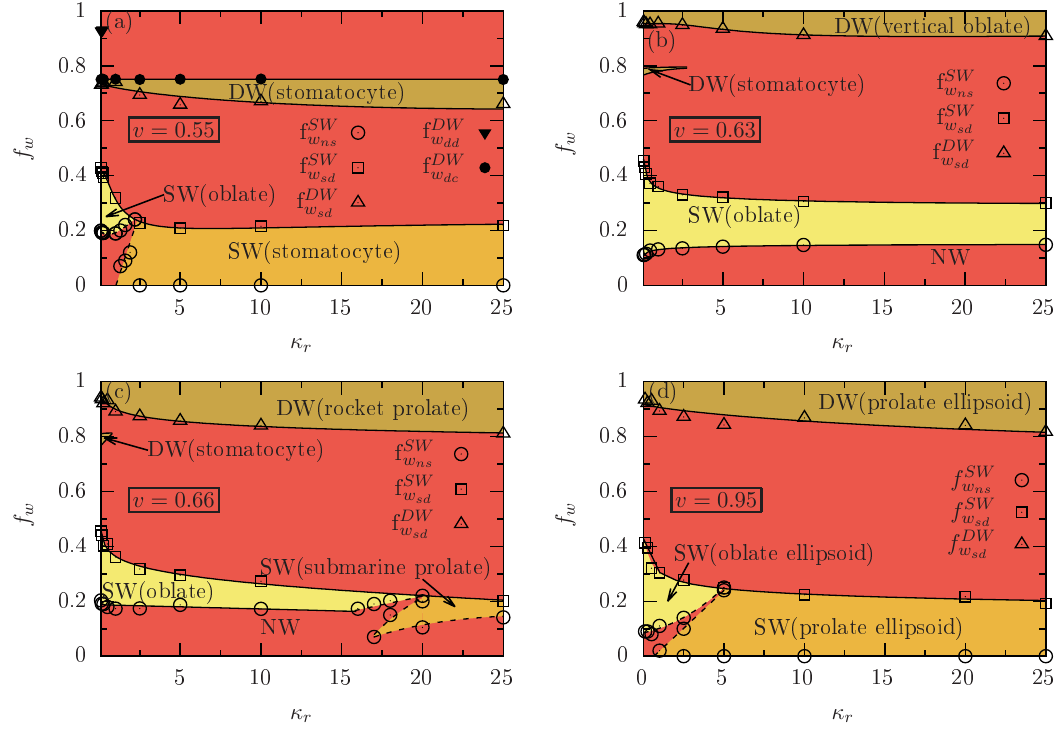}
    \caption{Wrapping diagrams in the $\kappa_{\rm r}$-$f_{\rm w}$ at $\tilde{\sigma} = 0.5$ indicating stable non-wrapped (NW), shallow-wrapped (SW), deep-wrapped (DW), and complete-wrapped (CW) states and energy barriers. Data is shown for nonspherical vesicles with reduced volumes (a) $v = 0.55$ (stomatocyte), (b) $v = 0.63$ (oblate), (c) $v = 0.66$ (prolate), (d) $v = 0.95$ (prolate). The red-colored regions indicate wrapping fractions $f_{\rm w}$ for states on the energy barriers.}
    \label{fig:phasediagram_fw_vs_kr}
\end{figure*}
\begin{figure*}[hbt!]
    \centering
    \includegraphics[width=2.0\columnwidth]{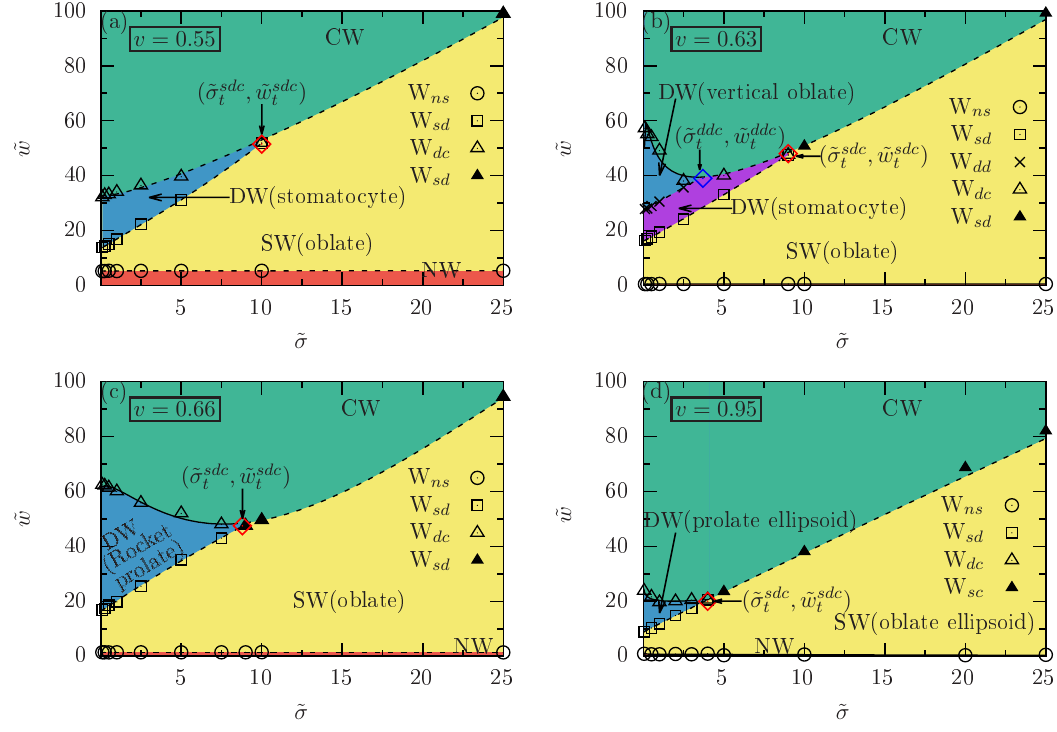}
    \caption{Wrapping diagrams in the $\tilde{\sigma}$-$\tilde{w}$ plane at $\kappa_{\rm r} = 1$. Data is shown for nonspherical vesicles with reduced volumes (a) $v = 0.55$ (stomatocyte), (b) $v = 0.63$ (oblate), (c) $v = 0.66$ (prolate), (d) $v = 0.95$ (prolate). Stable states and transitions are indicated analogously to Fig.~\ref{fig:phasediagram_w_vs_kr}; solid lines represent continuous and dashed lines discontinuous transitions, and triple points are highlighted by red diamonds.}
    \label{fig:phasediagram_w_vs_sigbar}
\end{figure*}
\subsection{\label{sec:shapes} Vesicle shapes during wrapping}
The shape of a particle can be quantified using the asphericity \cite{doi:10.1021/ma00148a028}, calculated from the principal moments ($\lambda_x$, $\lambda_y$, $\lambda_z$) of the gyration tensor,
\begin{equation}
    s^2=\frac{3}{2}\frac{\lambda_x^4+\lambda_y^4+\lambda_z^4}{(\lambda_x^2+\lambda_y^2+\lambda_z^2)^2}-\frac{1}{2}
\label{eq:asphericity}
\end{equation}
For a perfectly spherical particle, the principal moments are equal, $\lambda_x = \lambda_y = \lambda_z$, resulting in asphericity $s^2=0$. In contrast, the maximal asphericity $s^2=1$ is reached for rod-shaped particles, where $\lambda_x = \lambda_y = 0$ and $\lambda_z \neq 0$ when aligned along the z-axis. During wrapping, the soft vesicles exhibit various shapes depending on the reduced vesicle volume $v$ and the membrane tension $\tilde{\sigma}$. Thus, for nonspherical vesicles, $s^2$ ($> 0$) increases with increasing vesicle elongation. 

Figure \ref{fig:aspericity}(a) illustrates the asphericity $s^2$ of the vesicles with various reduced volumes $v$ as a function of the wrapping fraction $f_{\rm w}$ for $\kappa_{\rm r}=1$ and $\tilde{\sigma}=0.5$. Initially stomatocyte vesicles ($v=0.55$), resembling two connected spheres, have an asphericity close to zero. In the range $0 < f_{\rm w} \lesssim 0.07$, the vesicles retain the stomatocytic shape and $s^2\approx 0$. In stable SW states with $0.07 \lesssim f_{\rm w} \lesssim 0.5$, the vesicles are oblate with asphericity $s^2\simeq 0.23$. In the DW state, the vesicles undergo a sequence of shape transitions from stomatocytes to vertical oblates, and then back to inverted stomatocytes. This shape transition leads to a variation in asphericity from $s^2 \approx 0$ in the range $0.5 \lesssim f_{\rm w} \lesssim 0.75$, to $s^2 \simeq 0.23$ for $0.75 \lesssim f_{\rm w} \lesssim 0.9$, and finally back to $s^2\approx 0$ for $f_{\rm w} \gtrsim 0.9$.  

Initially oblate vesicles ($v=0.63$) retain constant asphericity $s^2 \simeq 0.23$ in the SW state for $0< f_{\rm w} \lesssim 0.5$, as they retain their shapes, see Fig.~\ref{fig:aspericity}(a). However, they transition to deep-wrapped stomatocytes for $0.5 \lesssim f_{\rm w} \lesssim 0.79$, reducing asphericity to $s^2 \simeq 0$. Within the DW state, the vesicles undergo a transition from stomatocyte to vertical oblate, returning the asphericity to $s^2\simeq 0.23$ for $0.79 < f_{\rm w} \leq 1$. In contrast, initially prolate vesicles ($v\simeq 0.66$), which are more elongated, have a high asphericity $s^2\simeq 0.97$. In SW states, $0.07 \lesssim f_{\rm w} \lesssim 0.5$, they transition to oblates with $s^2\simeq 0.23$. In DW states, $0.5 \lesssim f_{\rm w} \lesssim 0.81$, vesicles adopt stomatocytic states with $s^2\approx 0$. Finally, for $f_{\rm w} > 0.81$, the vesicles return to prolates with a rotation of the major axis by 90\textdegree and asphericities $s^2\simeq 0.97$. For ellipsoidal vesicles ($v=0.95$), the asphericity takes a value of $s^2\simeq 0.1$ when the vesicle adopts an oblate-ellipsoid shape in SW states for $0.1<f_{\rm w}\leq0.5$. In DW states, the vesicles maintain prolate-ellipsoidal shape for $0.5<f_{\rm w}\leq1$ with $s^2\simeq 0.38$.

Figure~\ref{fig:aspericity}(b) illustrates the effects of the stiffness of the vesicle, $\kappa_{\rm r}$, on the asphericity $s^2$ for an initially oblate vesicle ($v=0.63$) at membrane tension $\tilde{\sigma}=0.5$. For soft vesicles ($\kappa_{\rm r} \lesssim 2.5$), both shape and orientation changes occur during the wrapping process, resulting in variations in asphericity. Across all values of $\kappa_{\rm r}$, the vesicles retain their oblate shape in the SW state for $0<f_{\rm w} \lesssim 0.5$, maintaining an asphericity $s^2 \simeq 0.23$. However, in the DW state, the soft vesicles undergo a sequence of shape transitions as the wrapping fraction increases. For example, vesicles with $\kappa_{\rm r}=0.1$ adopt stomatocytic shapes with $s^2 \approx 0$ for $0.5 \lesssim f_{\rm w}\lesssim 0.72$, prolate shapes with $s^2\approx 0.97$ for $0.72 \lesssim f_{\rm w} \lesssim 0.9$, and vertical oblate shapes with $s^2\simeq 0.23$ for $f_{\rm w} \gtrsim 0.95$. In contrast, stiff vesicles ($\kappa_{\rm r}=10$) undergo only orientation changes, maintaining asphericity $s^2\simeq 0.23$. Similarly, the effects of membrane tension $\tilde{\sigma}$ on the asphericity of the vesicle are shown in Fig.~\ref{fig:aspericity}(c) for an oblate ($v=0.63$) with relative stiffness $\kappa_{\rm r}=1$ at various membrane tensions. Across a wide range of wrapping fractions $f_{\rm w}$ the asphericity $s^2$ remains nearly constant and is independent of $\tilde{\sigma}$, indicating that the shape of the vesicles is not much affected by changes in membrane tension. It decreases to $s^2 \lesssim 0.1$ for $0.6 \lesssim f_{\rm w} \lesssim 0.8$, where the vesicle is a stomatocyte. For initially stomatocyte, prolate, and ellipsoidal vesicles, the effect of vesicle stiffness and membrane tension on asphericities are shown in the SI; see Fig.~S6. The height of the center of mass of the vesicles with respect to the outer patch of the planar membrane can similarly indicate both shape change and orientation change during wrapping; see Fig.~S7 in the SI.

\subsection{\label{sec:vesstiffness} Effect of vesicle stiffness on the wrapping diagrams} 
Figures~\ref{fig:phasediagram_w_vs_kr}(a)-(d) present the wrapping diagrams for various stiffnesses $\kappa_{\rm r}$ and reduced adhesion strengths $\tilde{w}$ for vesicles with $v = 0.55$, $0.63$, $0.66$, and $0.95$ at fixed membrane tension $\tilde{\sigma} = 0.5$. These wrapping diagrams were extracted from the variation of the deformation energies, $\Delta \tilde{E}$ with $f_{\rm w}$; see Fig.~S4 in the SI. For each value of $v$, the binding transition $W_{\rm ns}$ shifts to higher adhesion strengths $\tilde{w}$ as $\kappa_{\rm r}$ increases, ultimately saturating at a constant value as the deformation of the vesicles is negligible for $\kappa_{\rm r} \gtrsim 10$. This implies that softer vesicles adhere to the membrane more easily than stiffer ones. In contrast, $\tilde{w}$ for the envelopment transition $W_{\rm dc}$ increases with decreasing $\kappa_{\rm r}$ due to the development of a sharp neck at the contact line; see Figs.~\ref{fig:snapshots} and ~\ref{fig:aspericity}, indicating that complete wrapping is more challenging for softer vesicles than for stiffer ones. The shape and stiffness of the vesicles also influence the nature of the binding transition, which can be continuous or discontinuous. For example, the binding transition is continuous for ellipsoidal vesicles ($v=0.95$) with $\kappa_{r} \gtrsim 1$ and stiff ($\kappa_{r} \gtrsim 1$) stomatocytes ($v=0.55$) vesicles, while it is discontinuous for soft ($\kappa_{\rm r} \lesssim 1 $) ellipsoidal ($v=0.95$), and soft ($\kappa_{\rm r} \lesssim 2.5 $) stomatocyte ($v=0.55$) vesicles. For oblate ($v=0.63$) and prolate ($v=0.66$) vesicles, the binding transition is discontinuous for the entire range of $\kappa_{\rm r}$. The envelopment transition is discontinuous for stomatocytes and continuous for oblates and prolates. The shallow-to-deep-wrapped transition $W_{\rm sd}$ is always discontinuous and shows a moderate increase of $\tilde{w}$ as $\kappa_{\rm r}$ decreases.

In the SW state, soft stomatocytes ($v = 0.55$) with bending rigidity ratio $\kappa_{\rm r} \lesssim 1$ adopt oblate shapes, while stiff ones with $\kappa_{\rm r} \gtrsim 2.5$ retain their original shape. For intermediate stiffness, $1 \lesssim \kappa_{\rm r} \lesssim 2.5$, a discontinuous transition from stomatocyte to oblate occurs as the adhesion strength $\tilde{w}$ increases; see Fig.~\ref{fig:phasediagram_w_vs_kr}(a). In contrast, oblate ($v = 0.63$) vesicles consistently maintain a "rocket" orientation for all bending-rigidity ratios; see Fig.~\ref{fig:phasediagram_w_vs_kr}(b). In the DW state, soft stomatocytes with $\kappa_{\rm r} < 1$ transition to vertical oblates with increasing $\tilde{w}$. While soft oblate vesicles ($v = 0.63$) undergo a shape transition as $\tilde{w}$ increases, moving from stomatocyte to rocket prolate for $\kappa_{\rm r} \lesssim 0.5$ and from stomatocyte to vertical oblate for $0.5 < \kappa_{\rm r} \lesssim 2.5$. Stiffer vesicles mainly experience an orientation change rather than a shape change. In complete wrapping, regardless of the bending-rigidity ratio, the vesicles return to their initial shapes. 

Initially prolate vesicles ($v = 0.66$) with $\kappa_{\rm r} \lesssim 16$ transition to oblate in the SW state, while those with $\kappa_{\rm r} \gtrsim 20$ maintain their shape, see Fig.~\ref{fig:phasediagram_w_vs_kr}(c). For intermediate stiffness values, $16 \lesssim \kappa_{\rm r} \lesssim 20$, the vesicles undergo a transition from prolate to oblate with increasing $\tilde{w}$. In the DW state, the vesicles with $\kappa_{\rm r} < 1$ experience a discontinuous transition from stomatocyte to "rocket" prolate with increasing $\tilde{w}$. In contrast, stiff prolates ($\kappa_{\rm r} > 1$) show only an orientation change, maintaining the "rocket" prolate state. Similarly, initially ellipsoidal vesicles ($v = 0.95$) with $1 \lesssim \kappa_{\rm r} \lesssim 5$ can exhibit a discontinuous transition from prolate to oblate in the SW state as $\tilde{w}$ increases; see Fig.~\ref{fig:phasediagram_w_vs_kr}(d). Softer shallow-wrapped vesicles ($\kappa_{\rm r} \lesssim 1$) remain in the oblate state throughout, while stiffer ones ($\kappa_{\rm r} \gtrsim 5$) remain in the prolate state. In the DW state, all ellipsoidal vesicles adopt the prolate shape with a "rocket" orientation.

Figure~\ref{fig:phasediagram_fw_vs_kr} shows the wrapping diagrams in the $\kappa_{\rm r}$–$f_{\rm w}$ plane for the same parameter values as in Fig.~\ref{fig:phasediagram_w_vs_kr}. For stomatocytes ($v = 0.55$) and ellipsoidal ($v = 0.95$) vesicles with $\kappa_{\rm r} \gtrsim 1$, initial binding to the planar membranes occurs at $f_{\rm w_{ns}} = 0$, indicating that the binding transition is continuous. In contrast, soft ($\kappa_{\rm r} \lesssim 1 $) ellipsoidal ($v=0.95$), soft ($\kappa_{\rm r} \lesssim 1 $) stomatocyte ($v=0.55$), prolate ($v = 0.66$), and oblate ($v = 0.63$) vesicles show discontinuous binding transitions to $f_{\rm w_{ns}} > 0$. For all reduced volumes, the wrapping fractions $f_{\rm w_{sd}}^{\rm SW}$ and $f_{\rm w_{sd}}^{\rm DW}$ for the coexising SW to DW states decrease weakly with increasing $\kappa_{\rm r}$ and may saturate for higher $\kappa_{\rm r}$. For stomatocyte ($v=0.55$), the wrapping fractions $f_{\rm w_{dc}}^{\rm DW}$ for the DW states that coexist with the CW state remain constant with increasing $\kappa_{\rm r}$. For initially stomatocytes ($v=0.55$), the presence of the energy barrier between the DW and CW states indicates that the envelopment transition is discontinuous. In contrast, for oblate ($v=0.63$), prolate ($v=0.66$), and ellipsoidal ($v=0.95$) vesicles, no such energy barrier is found, indicating that the envelopment transition is continuous. The range of accessible wrapping fractions for the DW states increases slightly with increasing $v$. 

\subsection{\label{sec:memtension} Effect of membrane tension on the wrapping diagrams}
The effects of membrane tension on the wrapping diagrams are shown in Fig.~\ref{fig:phasediagram_w_vs_sigbar} for stomatocyte ($v=0.55$), oblate ($v=0.63$), prolate ($v=0.66$) and ellipsoidal ($v=0.95$) vesicles, all with fixed relative stiffness $\kappa_{\rm r}=1$; the corresponding deformation energies are shown in Fig.~S5 in the SI. The adhesion strength $\tilde{w}$ for the SW-to-DW transition increases with increasing membrane tension $\tilde{\sigma}$, due to the increased energy cost in the neck, stabilizes SW over the DW states. As a result, the $W_{\rm dc}$ transition merges with the $W_{\rm sd}$ transition at the triple point ($\tilde{\sigma}^{\rm sdc}_t$, $\tilde{w}^{\rm sdc}_t$), where the SW, DW, and CW states coexist; the deformation energy landscapes at the triple points are shown in Fig.~S8 in the SI. Beyond the triple point, the envelopment transition becomes discontinuous, with vesicles undergoing a direct transition from the SW state to the CW state. The adhesion strength $\tilde{w}$ for the binding transition does not change with increasing membrane tension $\tilde{\sigma}$. 

For oblates ($v=0.63$) with a bending-rigidity ratio $\kappa_{\rm r}=1$, a deep-wrapped (${\rm DW}_1$) to deep-wrapped (${\rm DW}_2$) transition occurs at membrane tensions $\tilde{\sigma} \lesssim 4$, where stomatocytes and oblates coexist. For $\tilde{\sigma} \gtrsim 4$, vertical oblate states become unstable, and the ${\rm DW}_1$-to-${\rm DW}_2$ transition vanishes at another triple point ($\tilde{\sigma}^{\rm ddc}_t$, $\tilde{w}^{\rm ddc}_t$), where two distinct DW states coexist with the CW state; the corresponding deformation energy landscape is shown in Fig.~S9 in the SI.  For initially stomatocytes ($v=0.55$), the adhesion strength $\tilde{w}$ for complete wrapping gradually increases with increasing membrane tension $\tilde{\sigma}$, since the envelope transition remains discontinuous throughout the membrane tension range $\tilde{\sigma}$.  In contrast, for oblates ($v=0.63$), the envelopment transition is continuous for membrane tensions $0 < \tilde{\sigma} \lesssim \tilde{\sigma}^{\rm ddc}_t$ and discontinuous for $\tilde{\sigma} > \tilde{\sigma}^{\rm ddc}_t$. Similarly, for prolate and ellipsoidal vesicles, the envelopment transition is continuous for membrane tensions $0 < \tilde{\sigma} \lesssim \tilde{\sigma}^{\rm sdc}_t$ and discontinuous for $\tilde{\sigma} \gtrsim \tilde{\sigma}^{\rm sdc}_t$. Consequently, the adhesion strength $\tilde{w}$ for the complete wrapping initially decreases with increasing $\tilde{\sigma}$, reaches its optimal value at the triple point, and beyond the triple point $\tilde{w}$ increases with membrane tension $\tilde{\sigma}$. The wrapping diagram in the $f_{\rm w}-\tilde{\sigma}$ plane for initially stomatocyte, oblate, prolate, and ellipsoidal vesicles for the same parameter values, as mentioned in Fig.~\ref{fig:phasediagram_w_vs_sigbar}, is presented in Fig.~S10 in the SI.

\begin{figure}[h!]
    \centering
    \includegraphics[width=1.0\columnwidth]{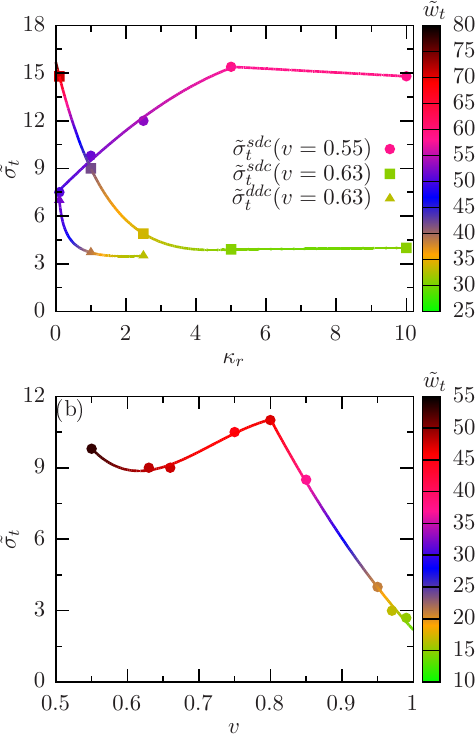}
    \caption{Membrane tensions $\tilde{\sigma}_t$ and adhesion strengths $\tilde{w}_t$ for the triple points as functions of (a) bending rigidity ratio $\kappa_{\rm r}$ for initially stomatocytic ($v = 0.55$, circles) and oblate ($v = 0.63$, squares and triangles) vesicles and (b) for vesicles with various reduces volumes $v$ at $\kappa_{\rm r} = 1$.} 
    \label{fig:Tpoint}
\end{figure}

The coordinates of the triple points in the $\tilde{w}$-$\tilde{\sigma}$ wrapping diagram are influenced by both the bending rigidity ratio $\kappa_{\rm r}$ and the reduced volume $v$ of the vesicles. For stomatocytes ($v=0.55$), the triple point ($\tilde{\sigma}^{\rm sdc}_t$, $\tilde{w}^{\rm sdc}_t$) shifts to higher values of membrane tension $\tilde{\sigma}$ and adhesion strength $\tilde{w}$ as the bending rigidity ratio $\kappa_{\rm r}$ increases; see Figs.~\ref{fig:Tpoint}(a) and Fig.~S11 in the SI. For high $\kappa_{\rm r}$, the almost spherical shape of the outer surface reduces the cost of wrapping energy for deep-wrapped stomatocytes, and therefore, the regions of stable DW states extend to lower adhesion strengths and higher membrane tensions $\tilde{\sigma}$ with increasing $\kappa_{\rm r}$. However, as the membrane tension $\tilde{\sigma}$ increases, independent of the vesicle shape, the energy cost associated with the neck for the DW states increases, increasing the stability of both the SW and the CW states compared to the DW state.  

For an oblate ($v = 0.63$) and $\kappa_{\rm r} \lesssim 3$, the vesicles transition between an oblate SW and a stomatocyte DW state with increasing adhesion strength, before further transition to oblate DW or CW states, see Fig.~\ref{fig:Tpoint}(a) and Fig.~S12 in the SI. However, unlike initial stomatocytes, increasing $\kappa_{\rm r}$ leads to decreasing tensions $\tilde{\sigma}$ and adhesion strengths $\tilde{w}$ for both triple points ($\tilde{\sigma}^{\rm sdd}_t$, $\tilde{w}^{\rm sdd}_t$) and ($\tilde{\sigma}^{\rm sdc}_t$, $\tilde{w}^{\rm sdc}_t$). Increasing the bending-rigidity ratio $\kappa_{\rm r}$ stabilizes the oblate states over the stomatocyte states and eventually leads to a direct transition between the oblate SW and the oblate DW or the CW states. The oblate DW state, due to its asymmetrical shape, lacks the advantages that a stomatocytic shape has, and stable states are only found for low tensions. 

The triple point ($\sigma^{\rm sdc}_{\rm t}, \tilde{w}^{\rm sdc}_{\rm t}$) varies nonmonotonically with reduced volume $v$ at a fixed stiffness of the vesicle $\kappa_{\rm r}=1$, as shown in Fig.~\ref{fig:Tpoint}(b). For $0.8 < v< 1$, the triple point moves to higher membrane tensions and adhesion strengths as $v$ decreases. This occurs because the vesicle becomes more elongated with decreasing $v$, increasing the stability of the deep-wrapped prolate states relative to the shallow-wrapped oblate states. In contrast, for $0.55<v<0.8$, the vesicles tend to adopt an oblate shape in the SW state and a stomatocyte shape in the DW state. Consequently, the triple point shifts to lower membrane tensions and adhesion strengths.

\section{\label{sec:conclusions} Conclusions and outlook}
We calculated the shapes, orientations, and energies for the wrapping of initially prolate, oblate, and stomatocytic vesicles at planar membranes. A systematic variation of reduced volumes, bending rigidity ratios, and membrane tensions leads to a rich phase behavior. Whereas shallow-wrapped states are often oblate, deep-wrapped stabilize stomatocytic shapes. Transitions between shallow- and deep-wrapped states are always discontinuous, whereas the binding and envelopment transitions can be either continuous or discontinuous.  Softer vesicles undergo major shape changes during wrapping, while stiffer ones show only orientation changes, and at high tensions, the transition can occur directly from SW to DW states. Stomatocytic shapes can, because of their near-spherical shape, stabilize deep-wrapped over shallow-wrapped states and also for initially oblate and prolate vesicles. For soft vesicles, the binding transition is shifted to lower adhesion strengths, and the envelopment transition to higher adhesion strengths, in agreement with previous studies \cite{doi:10.1021/acsnano.2c05801,yi2011cellular}. 

Increasing membrane tension destabilizes deep-wrapped states that vanish at a triple point, leading to the coexistence of shallow-, deep-, and complete-wrapped states. Vesicle asphericities and triple-point coordinates can be used to systematically characterize phase diagrams as functions of elastic parameters. With increasing stiffness, the triple points between the shallow-wrapped, deep-wrapped, and complete-wrapped states shift to lower tensions and adhesion strengths for initially oblate vesicles and to higher values for initially stomatocytic vesicles. For oblate, prolate and ellipsoidal vesicles, the optimal membrane tension for uptake is finite, indicating that the adhesion strength required for complete wrapping is not the lowest for vanishing tension, as expected for hard particles \cite{doi:10.1021/nl403949h,C3SM50351H,deserno2003wrapping}. Because cells can regulate their membrane tension \cite{sitarska2020pay,sheetz1996modulation}, non-spherical vesicles can favor endocytosis or shallow-wrapped states; the latter may eventually facilitate fusion and direct delivery of the vesicle volume to the cytosol \cite{kube2017fusogenic}.

Vesicles are a very versatile class of elastic particles, exhibiting unique behaviors not observed in particles with 3D elasticity, such as polymeric particles. In particular, in the regime of stable deep-wrapped states, vesicles show internal discontinuous shape transitions with stomatocytic states extending to higher membrane tensions. Studying initial vesicle shapes close to the phase boundaries between stomatocytic, oblate, and prolate vesicles leads to an intriguing qualitatively novel finding compared to prolate vesicles with reduced volumes $v>0.7$, which we studied earlier \cite{doi:10.1021/acsnano.2c05801}. Our findings on the wrapping of nonspherical vesicles at planar membranes provide valuable insights for designing tailored deformable particles, with potential applications in membrane attachment, fusion, and endocytosis. Insights from this study may also advance biomedical research, particularly targeted drug delivery using deformable nanoparticles.

\section*{\bf Acknowledgements}
J.M. thanks to Thorsten Auth (Jülich) for valuable discussions and input on the wrapping behavior of nonspherical vesicles on planar membranes. J.M. acknowledges funding support from the Science and Engineering Research Board (SERB/ANRF) under the MATRICS grant (Grant No. MTR/2023/001538) and IIT Bhubaneswar. A.K.S. thanks the Council of Scientific and Industrial Research (CSIR) for funding his PhD.

\section*{\bf Author Contributions}
JM designed the research. AKS, RM, and JM conducted the research and analyzed the data. JM and AKS drafted the manuscript.
\section*{\bf Conflicts of interest}
There are no conflicts to declare.

\section*{\bf Data availability}
The data used in this study are available in the main manuscript or the Electronic Supplementary Information (ESI). Additional raw data and analysis scripts can be obtained from the corresponding authors upon request.
\bibliography{reference} 

\end{document}


\preprint{APS/123-QED}

\title{Supporting Information:\\ Wrapping nonspherical vesicles at bio-membranes}

\author{Ajit Kumar Sahu} 
\affiliation{Department of Physics, School of Basic Sciences, Indian Institute of Technology Bhubaneswar, Jatni, Odisha-752050, India}%

\author{Rajkumar Malik}
\affiliation{Department of Physics, School of Basic Sciences, Indian Institute of Technology Bhubaneswar, Jatni, Odisha-752050, India}%

\author{Jiarul Midya}
\email{jmidya@iitbbs.ac.in}
\affiliation{Department of Physics, School of Basic Sciences, Indian Institute of Technology Bhubaneswar, Jatni, Odisha-752050, India}%

\maketitle

\section{Nonspherical vesicle shapes}
\begin{figure}[h]
    \centering
    \includegraphics[width=\columnwidth]{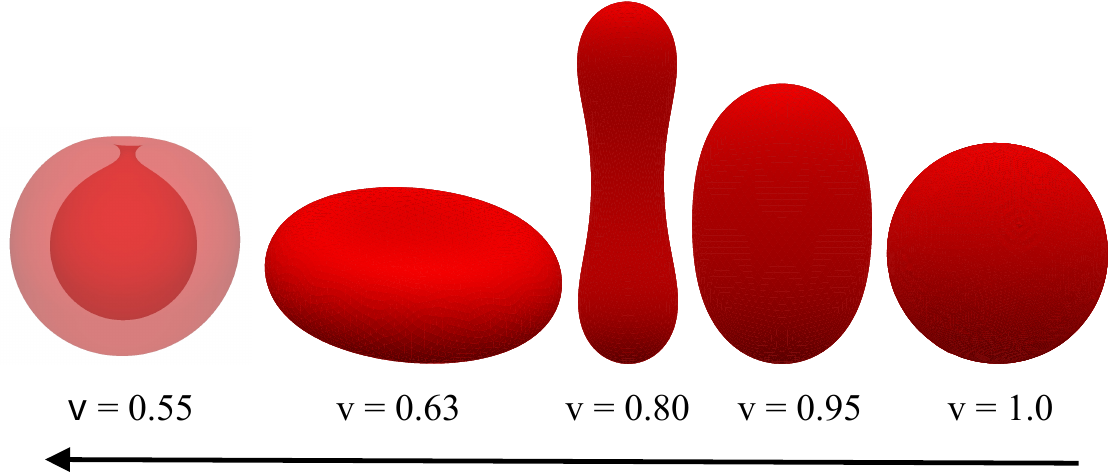}
    \caption{Nonspherical shapes of vesicle obtained by decreasing the reduced volume $v$, as indicated.}
    \label{figSI:ns_vesishape}
\end{figure}

Various nonspherical vesicles can be obtained by altering the volume of the vesicle while maintaining a constant surface area. Here we characterize the shapes of the vesicle by introducing a dimensionless parameter, reduced volume $v$, which is the ratio of the volume of the vesicle to that of a spherical vesicle with equal surface area. By decreasing the reduced volume, we can achieve different vesicle shapes; see Fig.~\ref{figSI:ns_vesishape}.

\begin{figure}[ht]
    \centering
    \includegraphics[width=\columnwidth]{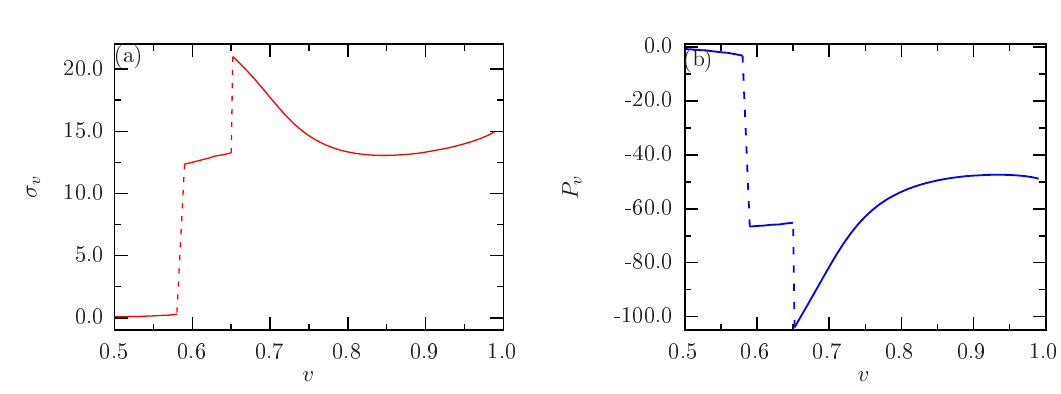}
    \caption{Variation of Lagrangian multipliers: (a) $\sigma_{\rm v}$, associated with area conservation, and (b) $P_{\rm v}$, associated with volume conservation, as functions of reduced volume $v$.}
    \label{figSI:Lagrange_multiplier}
\end{figure}

The shape change of the vesicle can also be analyzed using the Lagrangian multipliers $\sigma_{\rm v}$ and $P_{\rm v}$, which are associated with the area and volume of the vesicle membrane, respectively. Sharp changes in these values signify shape transitions. The value of $\sigma_{\rm v}$ monotonically decreases in the range $1 < v \lesssim 0.82$, it reaches a minimum at $v \simeq 0.82$, and then increases in the range $0.82 < v \lesssim 0.652$. In contrast $P_{\rm v}$ monotonically decreases in the range $1 < v \lesssim 0.652$, see Fig.~\ref{figSI:Lagrange_multiplier}(a) and (b), which correspond to a prolate shape. At $v\simeq 0.652$ a sharp change in $\sigma_{\rm v}$ and $P_{\rm v}$ is observed, indicating a shape change of the vesicle from prolate to oblate. A monotonic decrease in $\sigma_{\rm v}$ and $P_{\rm v}$ is observed in the range $0.652 < v \lesssim 0.592$, indicating that the vesicle remains oblate within this range of $v$.  Again, we observed a sudden decrease (increase) in the value of $\sigma_{\rm v}$ ($P_{\rm v}$) at $v=0.592$, indicating a change in the shape of the vesicle from oblate to stomatocyte. With further decrease in $v$, the values of $\sigma_{\rm v}$ and $P_{\rm v}$ remain almost constant, indicating that the shape of the vesicle remains stomatocyte in this range of $v$, as shown in Fig.~\ref{figSI:Lagrange_multiplier}(a) and (b).

\section{Triangulated membranes}
\begin{figure}[hbt!]
    \centering
    \includegraphics[scale=0.45]{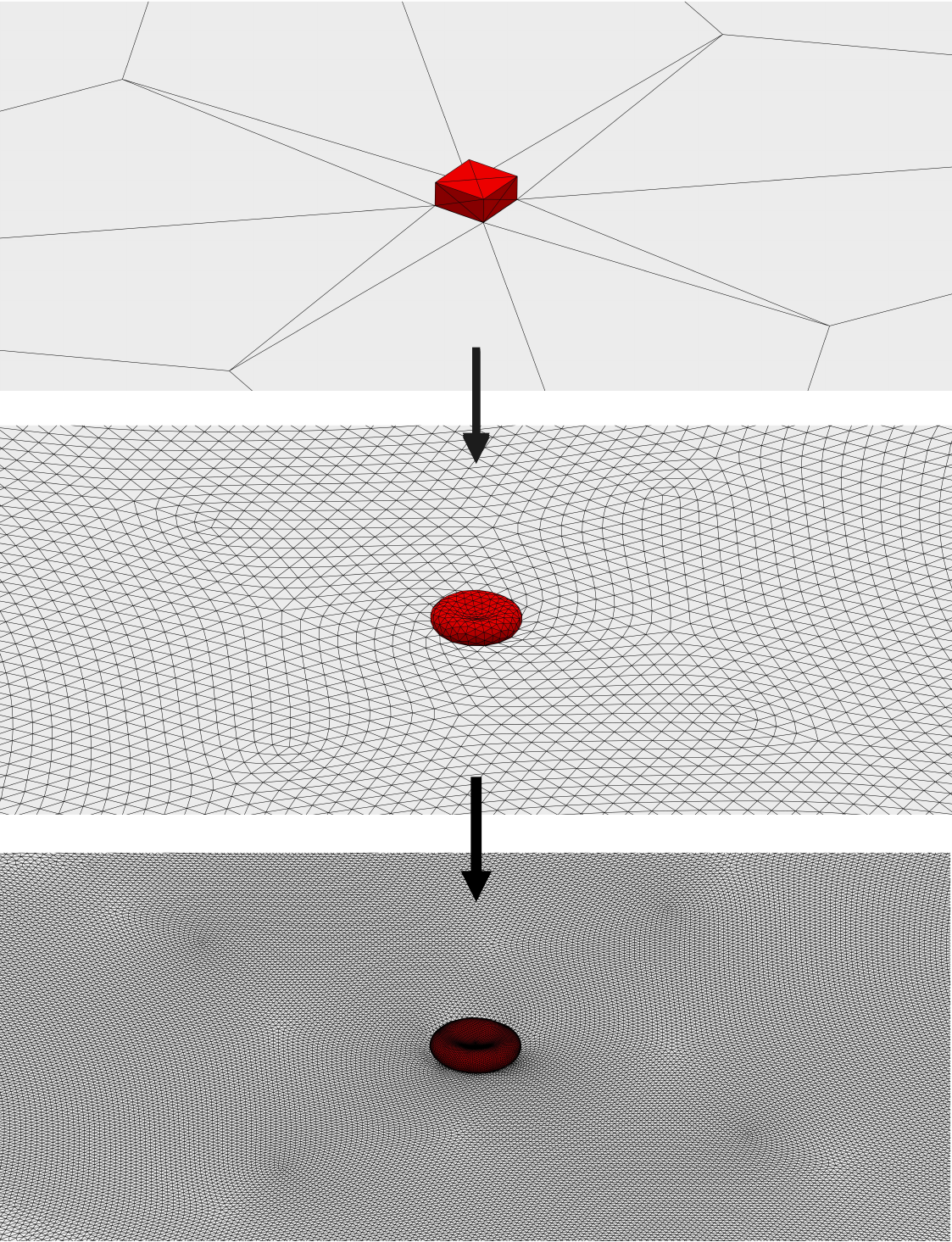}
    \caption{Discretized triangulated surfaces of the membrane-vesicle system at various stages of the refinement and energy minimization process.}
    \label{figSI:refineVesiMem}
\end{figure}

In our triangulated membrane model, the vesicle and membrane surfaces are discretized into triangles consisting of vertices, edges, and facets. The bending energy and surface energy were incorporated into the triangulated surface to behave it as a membrane. Constraints were applied to fix the volume and area of the vesicle membrane, while the area of the planar membrane was kept variable to allow the provision of additional membrane during wrapping, so that membrane tension remained constant. Initially, we started with a structure composed of larger triangles. To achieve the desired accuracy, the total energy of the system was minimized by refining the triangles. During this minimization process, triangle edges were also swapped to maintain the fluidity of the membrane. Figure~\ref{figSI:refineVesiMem} illustrates the different stages of refinement of the triangulated vesicle-membrane system.
\begin{figure}[hbt!]
    \centering
    \includegraphics[width=\figwidth]{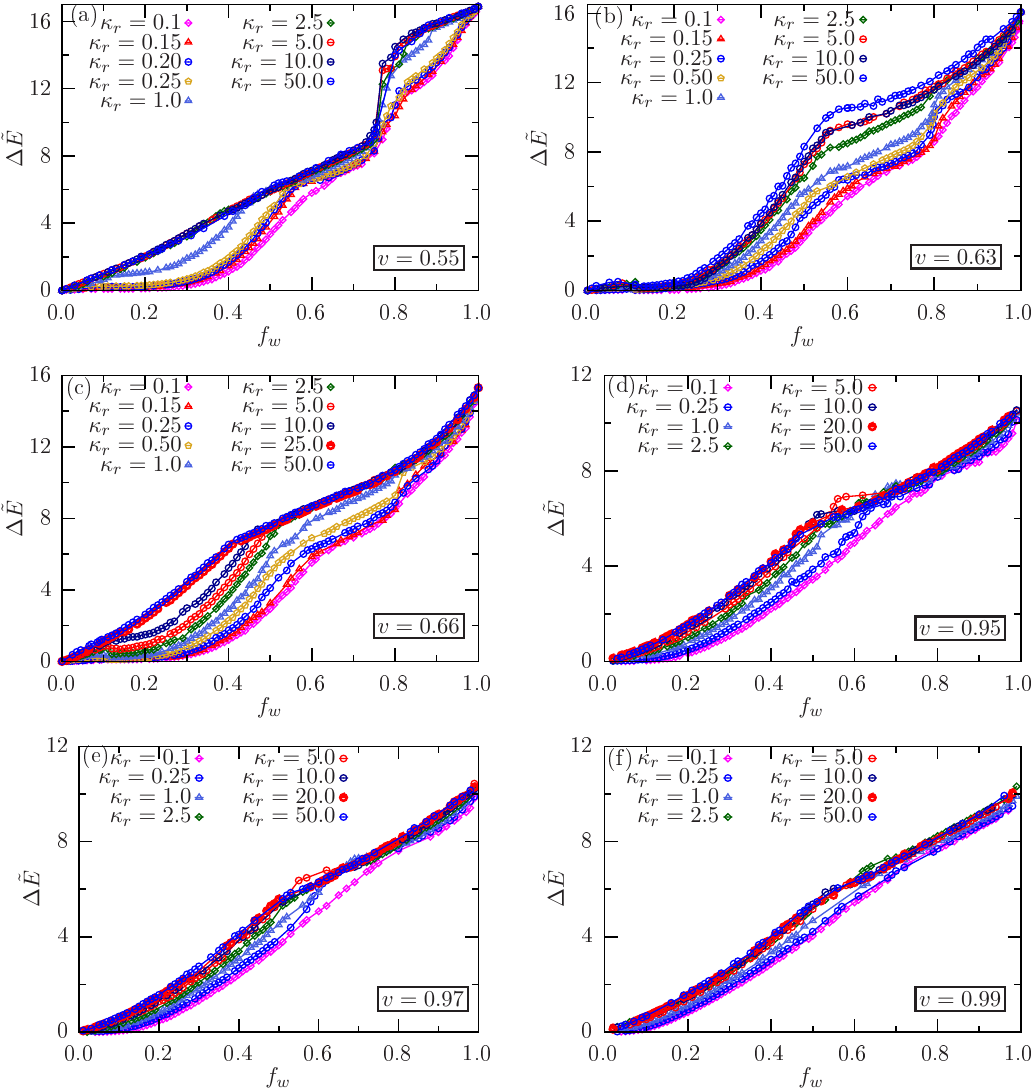}
    \caption{Reduced deformation energy $\Delta\tilde{E}$ as function of the wrapping fraction $f_{\rm {w}}$ for various values of $\kappa_{\rm {r}}$, at fixed $\tilde{\sigma}=0.5$ and reduced volume (a)  $v=0.55$, (b) $v=0.63$, (c) $v=0.66$, (d) $v=0.95$, (e) $v=0.97$, (f) $v=0.99$.}
    \label{figSI:defEnergy_Varying_kr}
\end{figure}
\section{Wrapping energy}
\begin{figure}[hbt!]
    \centering
    \includegraphics[width=\figwidth]{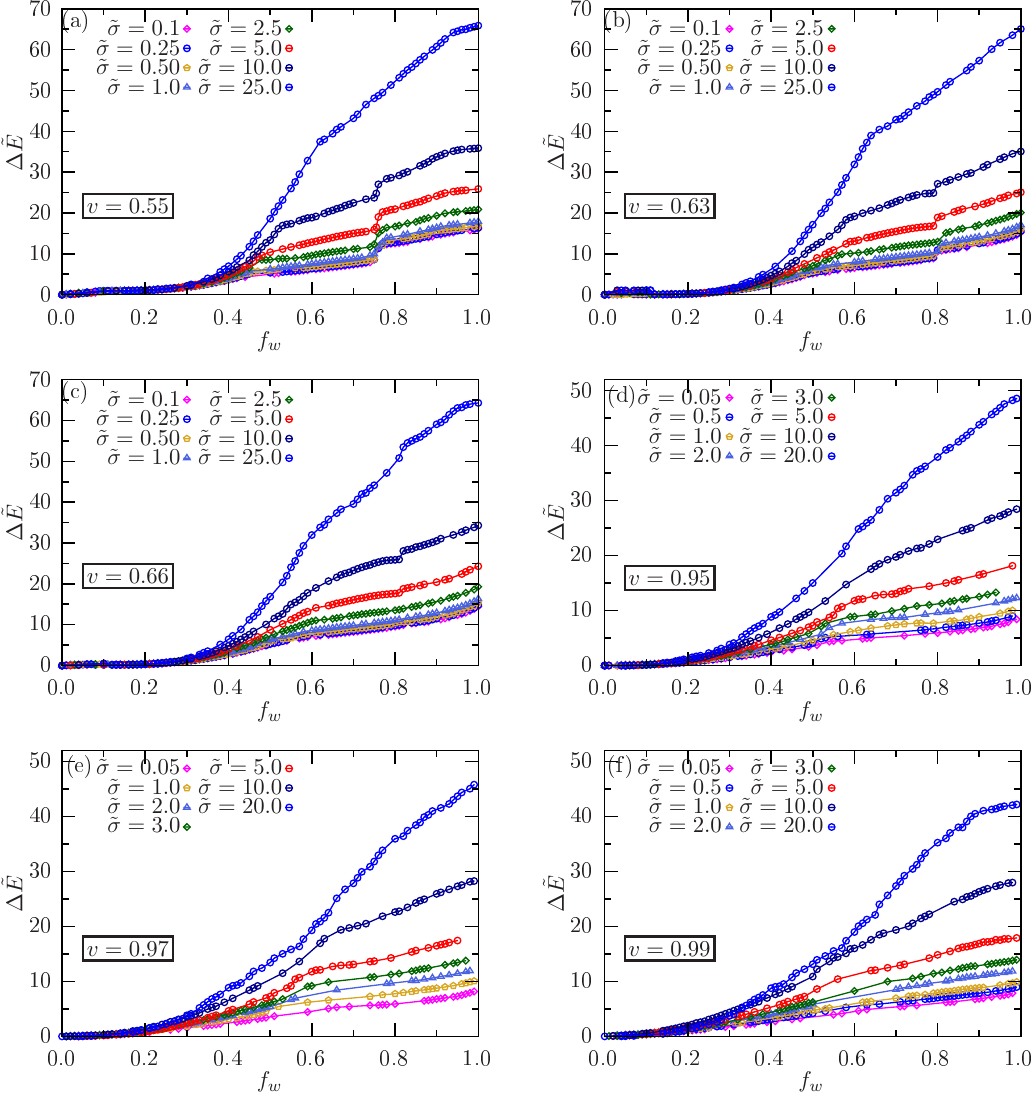}
    \caption{The deformation energy $\Delta\tilde{E}$ as function of the wrapping fraction $f_{\rm {w}}$ for various values of  $\tilde{\sigma}$, at fixed $\kappa_{\rm {r}} = 1.0$  and reduced volume (a) $v=0.55$, (b) $v=0.63$, (c) $v=0.66$, (d) $v=0.95$, (e) $v=0.97$, (f) $v=0.99$.}
    \label{figSI:E_vs_fw_sigBar}
\end{figure}

Using the Helfrich Hamiltonian, see Eq.(\ref{eq:Helfrich_Hamiltonian}) in the main text, we calculate reduced deformation energy, $\Delta{\tilde{E}}$ of the vesicle-membrane system as a function of wrapping fraction $f_{\rm{w}}$ for various reduced volumes $v$, and relative stiffness of the vesicle $\kappa_{\rm r}$ at fixed membrane tension $\sigma$; see Fig.~\ref{figSI:defEnergy_Varying_kr}. In our simulations, $\Delta{\tilde{E}}$ is calculated by fixing $f_{\rm w}$ to different values without applying any adhesion energy ($\tilde{w}=0$), and generating the deformation energy curve in the range of $f_{\rm w}$ [0: 1]. The results show that $\Delta{\tilde{E}}$ increases monotonically with $f_{\rm{w}}$, indicating that the non-wrapped state is energetically stable. For a given $v$, stiffer vesicles exhibit higher deformation energy than softer vesicles for $0 < f_{\rm{w}} < 1$, and the deformation energy increases as $v$ decreases at fixed $f_{\rm{w}}$. Similarly, Fig.~\ref{figSI:E_vs_fw_sigBar} illustrates $\Delta{\tilde{E}}$ for different membrane tensions $\tilde{\sigma}$ at fixed $\kappa_{\rm r}$. Without adhesion, $\Delta{\tilde{E}}$ increases with $f_{\rm{w}}$ and is higher for larger $\tilde{\sigma}$ at the same $f_{\rm{w}}$ and $v$, with the difference being more pronounced at higher $f_{\rm{w}}$. Furthermore, $\Delta{\tilde{E}}$ increases as $v$ decreases for a given $\tilde{\sigma}$.

\section{Quantification of vesicle shapes during wrapping}
\begin{figure}[hbt!]
    \centering
    \includegraphics[width=\figwidth]{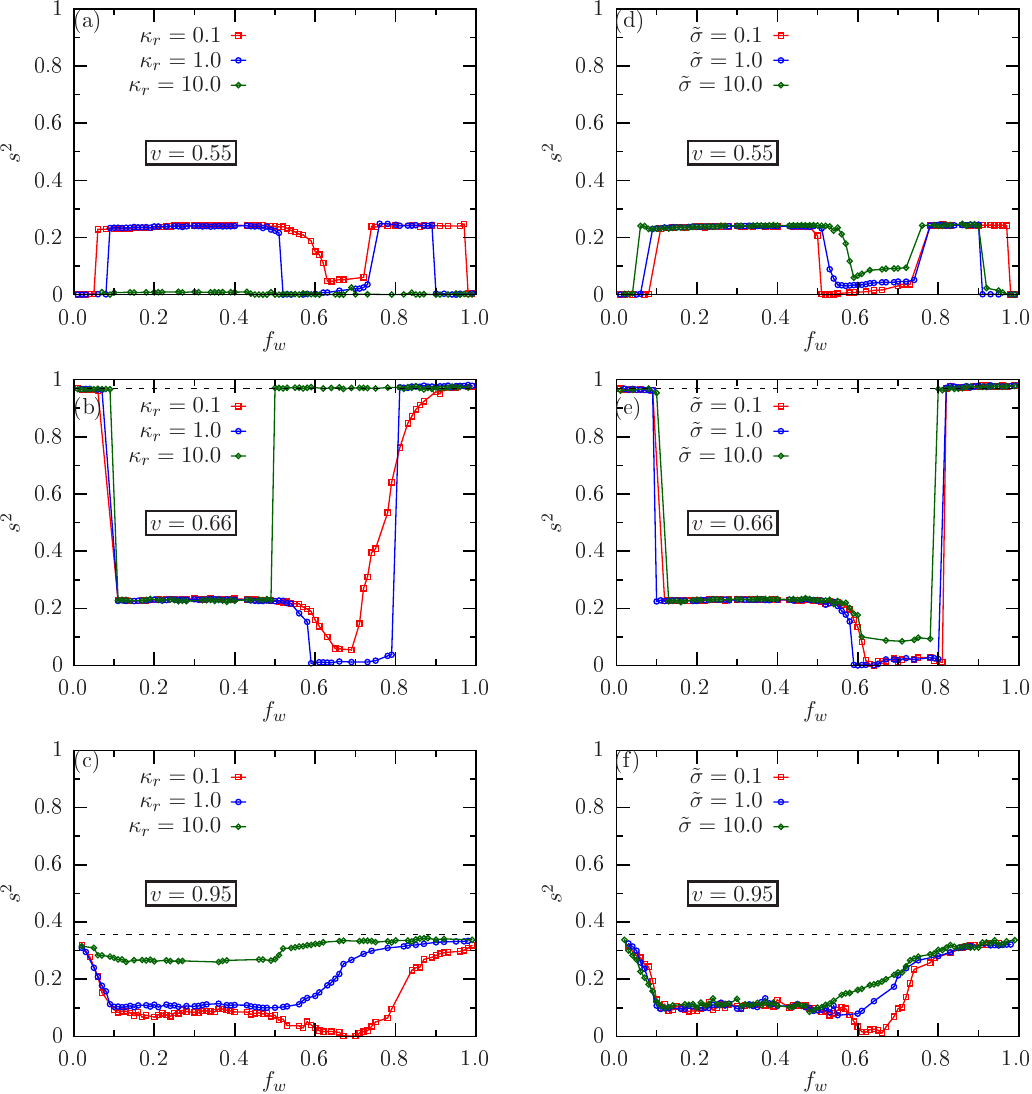}
    \caption{The asphericity $s^2$  of a vesicle, as a function of the wrapping fraction $ f_{\rm w}$ at fixed membrane tension $\tilde{\sigma} = 0.5$, and at reduce volume (a) $v = 0.55$, (b) $v = 0.66$, and (c) $v = 0.95$, for various vesicle stiffness $\kappa_{\rm r}$, as indicated.  The asphericity $s^2$ vs $ f_{\rm w}$ at fixed vesicle stiffness $\kappa_{\rm r} = 1$ and reduce volume (e) $v = 0.55$, (f) $v = 0.66$, and (g) $v = 0.95$, and for various membrane tension $\tilde{\sigma}$, as indicated. The dashed lines represent the asphericity of free vesicles with the corresponding reduced volumes.}
    \label{figSI:asphericity_all}
\end{figure}

\begin{figure}
    \centering
    \includegraphics[width=\figwidth]{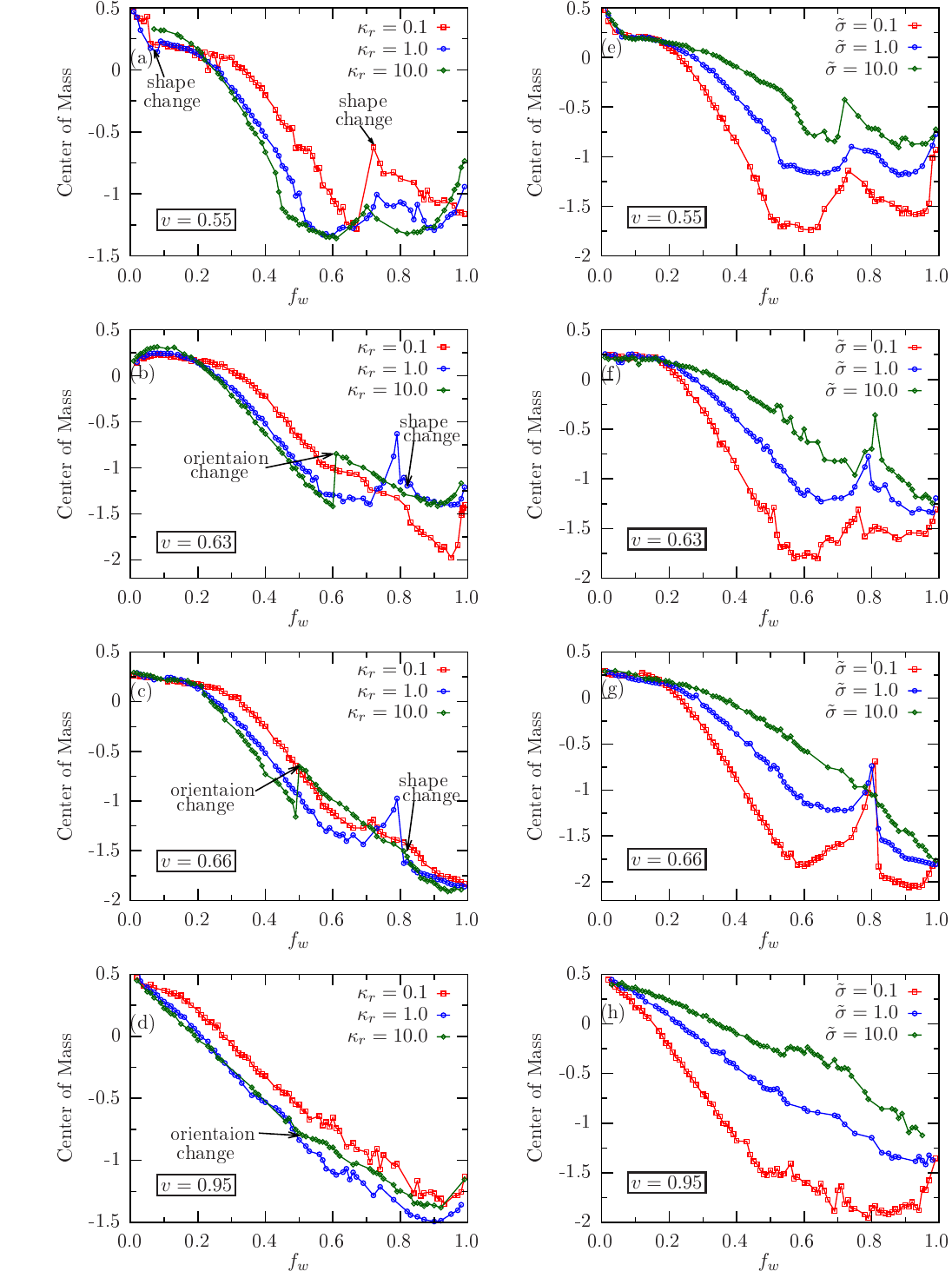}
    \caption{Variation of the relative height of the center-of-mass of the vesicle for different stiffness values of $\kappa_{\rm r}$ at fixed membrane tension $\tilde{\sigma} = 0.5$: (a) $v = 0.55$, (b) $v = 0.63$, (c) $v = 0.66$, (d) $v = 0.95$. Center of mass for different values of $\tilde{\sigma}$ at fixed stiffness $\kappa_{\rm r}=1$: (e) $v = 0.55$, (f) $v = 0.63$, (g) $v = 0.66$, (h) $v = 0.95$.}
    \label{figSI:CM_curve}
\end{figure}

The shape change of a vesicle during the wrapping process can be quantified by calculating its asphericity $s^2$ (see Eq.\ref{eq:asphericity} in the main text) and the center-of-mass relative to the height of the outer edge of the planar membrane. Figure\ref{figSI:asphericity_all} illustrates the variation of $s^2$ as a function of $f_w$ for different values of $v$, $\tilde{\sigma}$, and $\kappa_{\rm {r}}$. Any sudden change in $s^2$ indicates a change in the shape of the vesicle. By comparing the $s^2$ values of the wrapped vesicles with those of the free vesicles, we can infer shape changes under variation of stiffness and membrane tension. The relative height of the vesicle's center-of-mass can similarly indicate both shape and orientation changes during wrapping. Sudden changes in the relative center-of-mass height signal such transitions, see Fig.\ref{figSI:CM_curve}. For rigid spherical vesicles, the center of mass decreases monotonically with increasing wrapping fraction $f_{\rm w}$. In contrast, deformable vesicles exhibit nonmonotonic behavior as a result of shape and orientation changes. For example, initially stomatocyte ($v=0.55$) vesicles undergo a series of transitions: stomatocyte $\rightarrow$ oblate $\rightarrow$ stomatocyte $\rightarrow$ vertical oblate $\rightarrow$ inverted stomatocyte—resulting in sudden changes in center-of-mass height at the transition points, see Fig.\ref{figSI:CM_curve}(a). Similarly, oblate, prolate, and ellipsoidal vesicles show similar nonmonotonic behavior in center-of-mass height during wrapping.
\section{Triple point and wrapping diagrams in $\tilde{w}$-$\tilde{\sigma}$ plane}
\begin{figure}[hbt!]
    \centering
    \includegraphics[width=1.0\columnwidth]{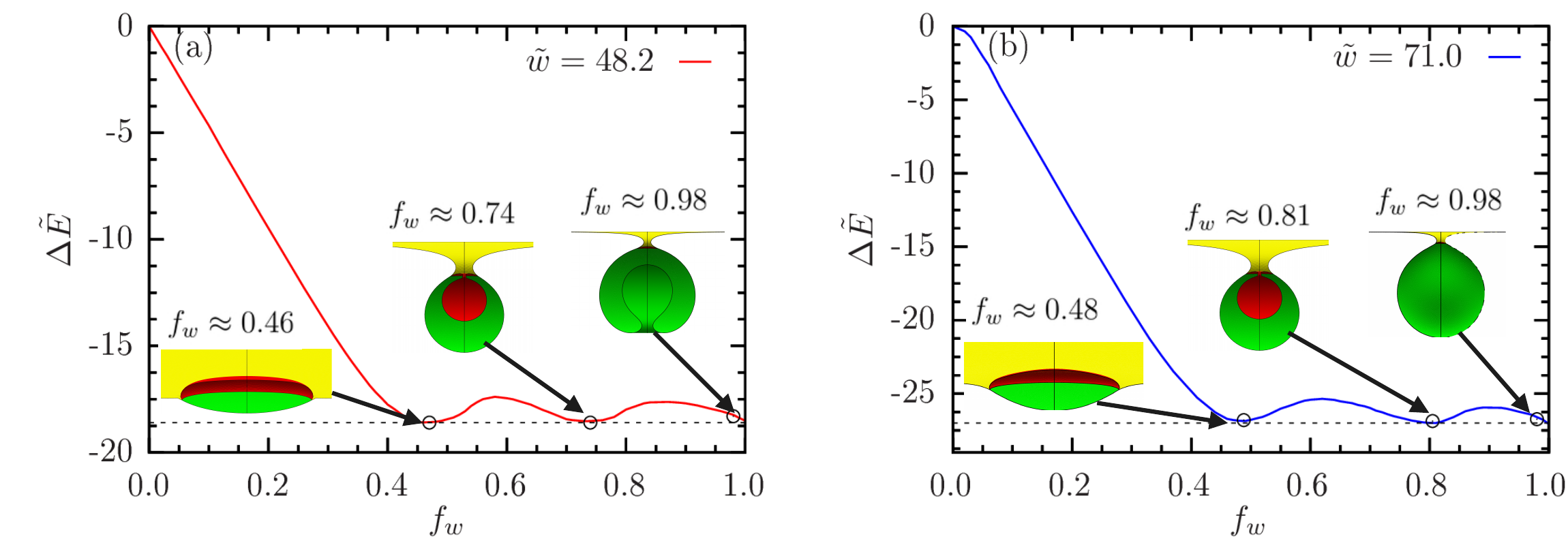}
    \caption{Deformation-energy landscapes for vesicles at the triple points for (a) initially stomatocytic vesicles at $v = 0.55$, $\tilde{\sigma} = 7.5$, and $\kappa_{\rm r} = 0.1$, and (b) initially oblate vesicles at $v = 0.63$, $\tilde{\sigma} = 14.8$, and $\kappa_{\rm r} = 0.1$.}
    \label{figSI:Tpoint_graph}
\end{figure}

\begin{figure}[hbt!]
    \centering
    \includegraphics[width=1.0\columnwidth]{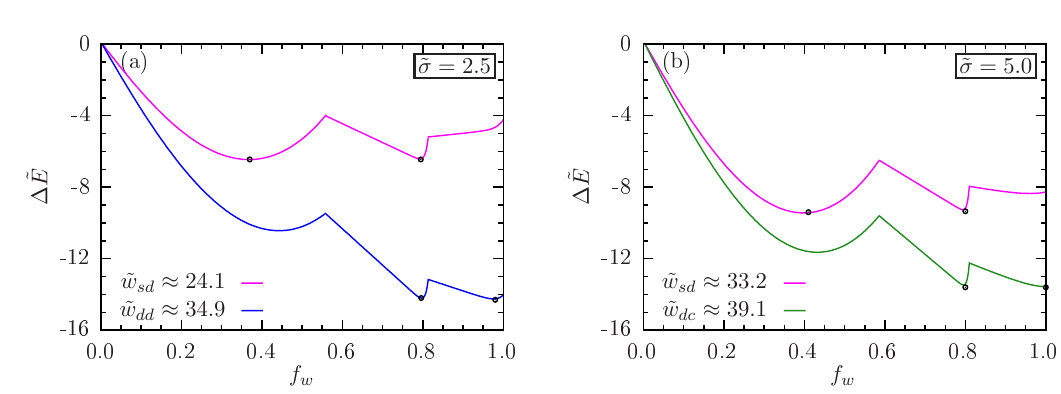}
    \caption{Deformation-energy landscapes for oblate vesicles ($v = 0.63$) with $\kappa_r = 1$: (a) shallow-to-deep wrap transition and deep-to-deep wrap transition at membrane tension $\tilde{\sigma} = 2.5$; (b) shallow-to-deep wrap transition and deep-to-complete-wrapped transition at membrane tension $\tilde{\sigma} = 5$. }
    \label{figSI:Tpoint_ddc}
\end{figure}

The triple point represents the coexistence of three stable states of the vesicle, where the shallow-wrapped (SW), deep-wrapped (DW), and complete-wrapped (CW) states coexist. Alternatively, two distinct deep-wrapped states may co-exist with the CW state, depending on reduced volume $v$, vesicle stiffness $\kappa_{\rm r}$, and membrane tension $\tilde{\sigma}$. In Fig.~\ref{figSI:Tpoint_graph}, we present the energy landscape for the initially stomatocyte ($v=0.55$) and oblate ($v=0.63$) vesicles at the triple point, where the stable SW, DW and CW states coexist at ($\tilde{\sigma}_t^{sdc}, \tilde{w}_t^{sdc}$). For oblate vesicles with $\kappa_{\rm r} \lesssim 3$, a second triple point is observed corresponding to the coexistence of two DW states with the CW state. Figure~\ref{figSI:Tpoint_ddc}(a) illustrates the SW-to-DW and DW-to-DW transitions for $\kappa_{\rm r} = 1$ at $\tilde{\sigma} = 2.5$. However, at $\tilde{\sigma} = 5$, the second deep-wrapped state becomes energetically unstable, and the first DW state coexists directly with the CW state; see Fig.~\ref{figSI:Tpoint_ddc}(b). Consequently, the DW-to-DW transition line merges with the DW-to-CW transition line, forming a triple point ($\tilde{\sigma}_t^{ddc}, \tilde{w}_t^{ddc}$) with stable DW, DW, and CW states; see Fig.~\ref{fig:phasediagram_w_vs_sigbar}(b) in the main text.

\begin{figure}[hbt!]
    \centering   
    \includegraphics[width=1.0\columnwidth]{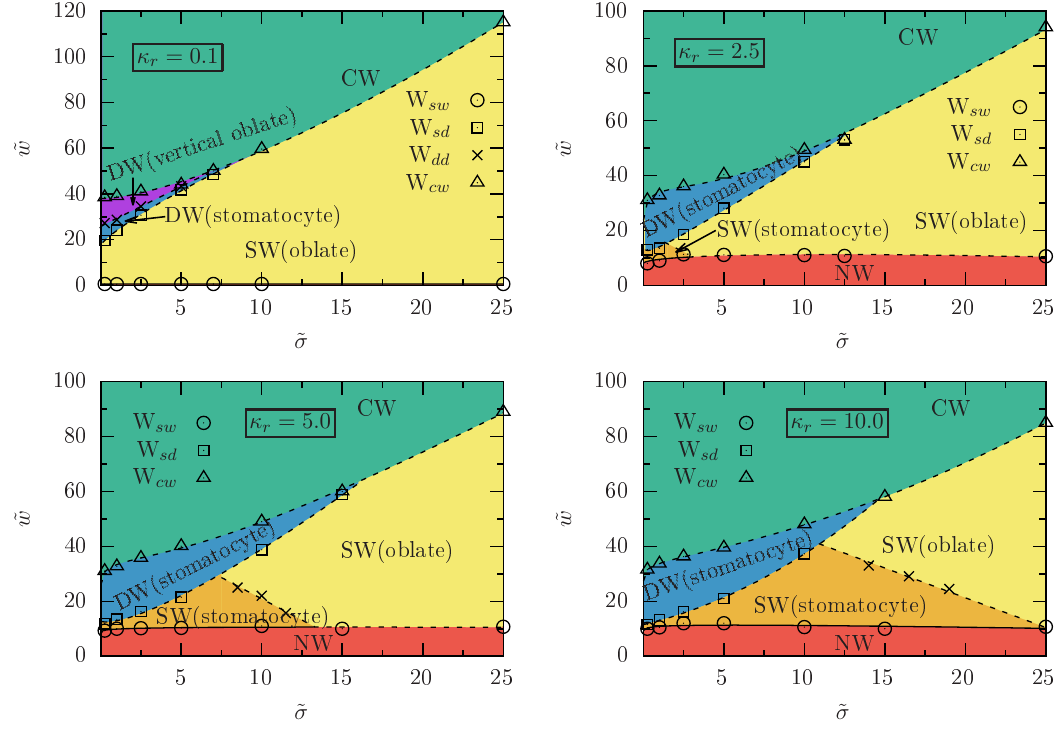}
    \caption{Wrapping diagrams for initially stomatocyte vesicles with $v = 0.55$ in the $\tilde{w}-\tilde{\sigma}$ plane for different stiffness values: (a) $\kappa_{\rm r} = 0.1$, (b) $\kappa_{\rm r} = 2.5$, (c) $\kappa_{\rm r} = 5$, and (d) $\kappa_{\rm r} = 10$. The different colors represent the stable wrapped states of the vesicles. Solid lines indicate continuous transitions, while dashed lines represent discontinuous transitions. }
    \label{figSI:wDia_stomatocyte}
\end{figure}

The stiffness of the vesicles affects the coordinates of the triple point. For initially stomatocytes ($v = 0.55$), the triple point ($\tilde{\sigma}_t^{sdc}, \tilde{w}t^{sdc}$) shifts to higher values of $\tilde{\sigma}$ and $\tilde{w}$ as $\kappa{\rm r}$ increases; see Fig.~\ref{figSI:wDia_stomatocyte}. This shift occurs due to the preference for the stomatocyte shape in the DW state, attributed to its spherical outer layer. Comparing the wrapping diagrams for initially stomatocytes ($v = 0.55$) with different stiffness values $\kappa_{\rm r}$, we find that the transition $W_{\rm cw}$ is nearly independent of $\kappa_{\rm r}$. However, the transition $W_{\rm sd}$ for a higher value of $\kappa_{\rm r}$ occurs at a lower $\tilde{w}$, suggesting that the stomatocyte becomes less energetically favorable in the SW state compared to the DW state as the stiffness of the vesicle increases.

\begin{figure}[hbt!]
    \centering
    \includegraphics[width=1.0\columnwidth]{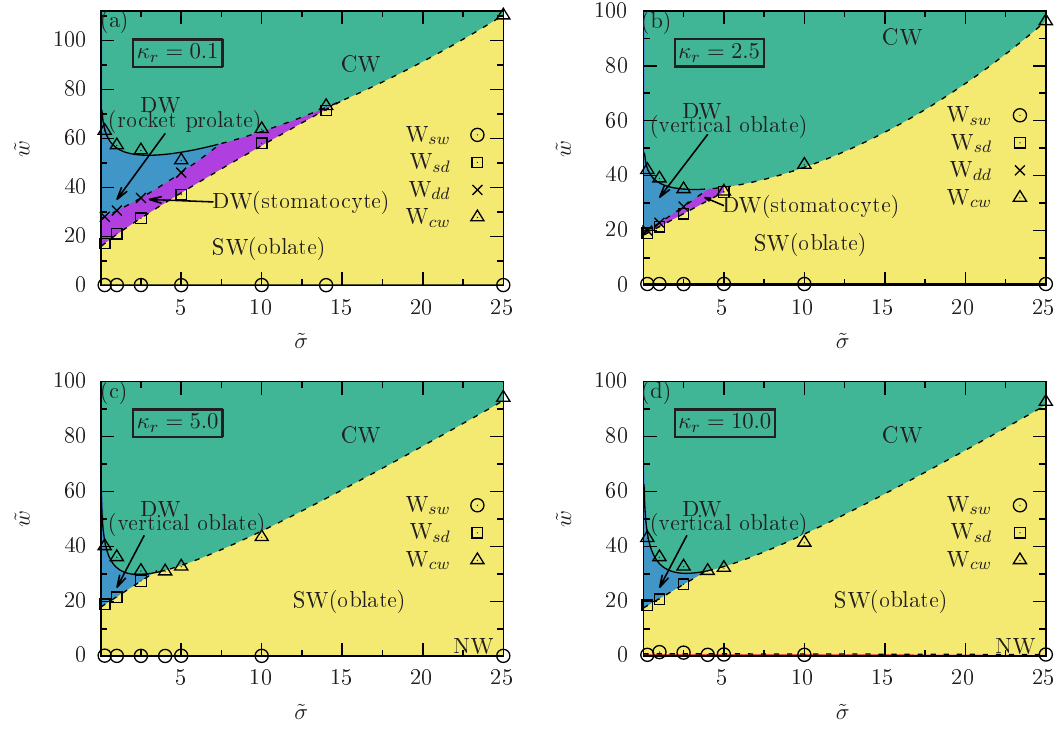}
    \caption{Wrapping diagrams for initially oblate vesicles with $v = 0.63$ in the $\tilde{w}-\tilde{\sigma}$ plane for different stiffness values: (a) $\kappa_{\rm r} = 0.1$, (b) $\kappa_{\rm r} = 2.5$, (c) $\kappa_{\rm r} = 5$, and (d) $\kappa_{\rm r} = 10$. The different colors represent the stable wrapped states of the vesicles. Solid lines indicate continuous transitions, while dashed lines represent discontinuous transitions.}
    \label{figSI:wDia_oblate}
\end{figure}

For oblate vesicles ($v=0.63$), both triple points shift to lower values of $\tilde{\sigma}$ and $\tilde{w}$ as the stiffness of the vesicle, $\kappa_{\rm r}$, increases; see Fig.~\ref{figSI:wDia_oblate}. In this case, the $W_{sd}$ transition remains unaffected by changes in vesicle stiffness. However, the $W_{cw}$ transition occurs at a relatively higher $\tilde{w}$ for softer vesicles. This indicates that the oblate shape is favored in the SW state, as it causes less deformation to the planar membrane in the shallow-wrapped configuration. In contrast, the oblate shape in the DW state induces a higher deformation energy in the system. Further, we observed that the optitmal adhesion strength for the complete wrapping always occurs at the triple point.

\section{Wrapping diagram in $f_{\rm {w}}$-$\tilde{\sigma}$ plane}
\begin{figure}[hbt!]
    \centering
    \includegraphics[width=1.0\columnwidth]{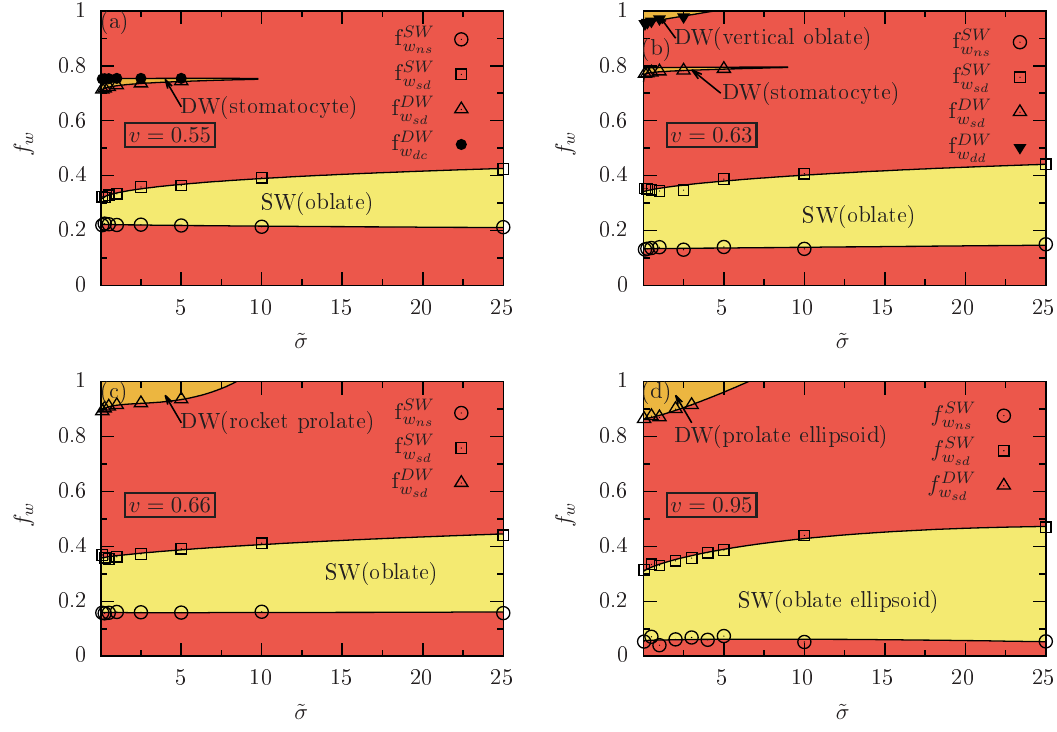}
    \caption{Wrapping diagram of non-spherical vesicles in $f_{\rm {w}}$-$\tilde{\sigma}$ plane for reduced volume (a) $v =0.55$ (stomatocyte), (b) $v = 0.63$ (oblate), (c) $v=0.66$ (prolate), (d) $v =0.95$ (ellipsoidal) at fixed $\kappa_{\rm {r}} = 1$. The red region shows the inaccessible wrapping fraction, and the yellow region shows the accessible wrapping
     fraction.}
    \label{figSI:wDia_fw_vs_sigBar}
\end{figure}

Figure~\ref{figSI:wDia_fw_vs_sigBar} illustrates the wrapping diagram in the $f_{\rm w}$-$\tilde{\sigma}$ plane for initially stomatocyte ($v=0.55$), oblate ($v=0.63$), prolate ($v=0.66$), and ellipsoidal ($v=0.95$) vesicles. The red region represents inaccessible values of $f_{\rm w}$ corresponding to unstable states, while the yellow region indicates stable states of the vesicles during the wrapping process. The stable shallow-wrapped state occurs at a non-zero value of $f_{\rm w}$ and remains constant with variations in membrane tension $\tilde{\sigma}$ for all reduced volumes $v$. For vesicles with $\kappa_{\rm {r}} = 1$, the oblate shape is consistently the stable shallow-wrapped configuration in the entire range of $\tilde{\sigma}$. The stability of the shallow-wrapped oblate state increases with increasing membrane tension $\tilde{\sigma}$ and reduced volume $v$. In contrast, the stable deep-wrapped state emerges only at lower values of $\tilde{\sigma}$ and vanishes as $\tilde{\sigma}$ increases. For $v=0.63$ and $v=0.55$, the stomatocyte is the stable deep-wrapped state. However, for ellipsoidal ($v=0.95$) and prolate ($v=0.66$) vesicles, the "rocket" prolate shape becomes stable in the deep-wrapped state. Interestingly, for $v=0.63$, two distinct stable deep-wrapped states are observed: one with a stomatocyte shape and the other with an oblate shape oriented perpendicularly to the membrane. This deep-to-deep wrapping transition occurs only under low membrane tension.

